\documentclass[sigconf]{acmart}
\makeatletter
\newcommand{\method}{\textit{MoRER}\@ifnextchar.{}{\@ifnextchar,{}{\@ifnextchar:{}{\@ifnextchar+{}{\@ifnextchar'{}{ }}}}}}
\makeatother
\makeatletter
\newcommand{\almser}{\textit{Almser}\@ifnextchar.{}{\@ifnextchar,{}{\@ifnextchar:{}{\@ifnextchar+{}{\@ifnextchar'{}{ }}}}}}
\makeatother
\makeatletter
\newcommand{\ditto}{\textit{Ditto}\@ifnextchar.{}{\@ifnextchar,{}{\@ifnextchar:{}{\@ifnextchar+{}{\@ifnextchar'{}{ }}}}}}
\makeatother
\makeatletter
\newcommand{\multiem}{\textit{MultiEM}\@ifnextchar.{}{\@ifnextchar,{}{\@ifnextchar:{}{\@ifnextchar+{}{\@ifnextchar'{}{ }}}}}}
\makeatother
\makeatletter
\newcommand{\sudowoodo}{\textit{Sudowoodo}\@ifnextchar.{}{\@ifnextchar,{}{\@ifnextchar:{}{\@ifnextchar+{}{\@ifnextchar'{}{ }}}}}}
\makeatother
\makeatletter
\newcommand{\transer}{\textit{TransER}\@ifnextchar.{}{\@ifnextchar,{}{\@ifnextchar:{}{\@ifnextchar+{}{\@ifnextchar'{}{ }}}}}}
\makeatother
\makeatletter
\newcommand{\unicorn}{\textit{Unicorn}\@ifnextchar.{}{\@ifnextchar,{}{\@ifnextchar:{}{\@ifnextchar+{}{\@ifnextchar'{}{ }}}}}}
\makeatother
\makeatletter
\newcommand{\any}{\textit{AnyMatch}\@ifnextchar.{}{\@ifnextchar,{}{\@ifnextchar:{}{\@ifnextchar+{}{\@ifnextchar'{}{ }}}}}}

\newcommand{\music}{\textit{Music}\@ifnextchar.{}{\@ifnextchar,{}{\@ifnextchar:{}{\@ifnextchar+{}{\@ifnextchar){}{ }}}}}}
\newcommand{\dexter}{\textit{Dexter}\@ifnextchar.{}{\@ifnextchar,{}{\@ifnextchar:{}{\@ifnextchar+{}{\@ifnextchar){}{ }}}}}}
\newcommand{\wdc}{\textit{WDC-computer}\@ifnextchar.{}{\@ifnextchar,{}{\@ifnextchar:{}{\@ifnextchar+{}{\@ifnextchar){}{ }}}}}}

\makeatother

\acmDOI{XXXXXXX.XXXXXXX}
\acmConference[EDBT 2026]{29th International Conference on Extending Database Technology (EDBT)}{24th March-27th March, 2026}{Tampere, Finland}
\acmYear{2026}
\usepackage{booktabs}
\usepackage{amsmath}
\usepackage{amsthm}
\newtheorem{definition}{Definition}
\usepackage{xcolor} 
\definecolor{forest}{rgb}{0.0, 0.27, 0.13}
\usepackage{graphicx}
\usepackage{csquotes}
\usepackage{enumitem}
\usepackage{subcaption}
\usepackage{multirow}
\usepackage{multicol}
\usepackage{multirow, tabularx, ragged2e}
\newcolumntype{C}{>{\Centering\arraybackslash}X}
\usepackage{colortbl}
\usepackage{lipsum}
\usepackage{dblfloatfix}
\usepackage{nicefrac}
\usepackage{tabularx}
\usepackage{cleveref}
\usepackage[nolist,nohyperlinks]{acronym}
\usepackage{xspace}
\usepackage{subfiles}
\crefname{algocf}{alg.}{algs.}
\Crefname{algocf}{Algorithm}{Algorithms}
\Crefname{equation}{Eq.}{Eqs.}
\Crefname{figure}{Fig.}{Figs.}
\Crefname{tabular}{Table}{Tables}
\crefname{table}{Table}{Tables}
\crefname{tabular}{Table}{Tables}
\Crefname{table}{Table}{Tables}
\Crefname{section}{Sect.}{Sects.}
\usepackage{array}
\usepackage{url}
\usepackage{pdfpages}
\usepackage[linesnumbered, ruled, vlined]{algorithm2e}

\SetAlFnt{\small}
\SetAlCapFnt{\small}
\SetAlCapNameFnt{\small}
\SetAlCapHSkip{0pt}
\IncMargin{-\parindent}
\AtBeginDocument{}

\begin{acronym}
    \acro  {er}   [ER]   {entity resolution}
    \acro {al} [AL] {active learning}
    \acro {ml} [ML] {machine learning}
     \acro {llm} [LLM] {large language model}
    \acrodefplural{llm}[LLMs]{large language models}
\end{acronym}

\begin{document}
\title{Efficient Model Repository for Entity Resolution: Construction, Search, and Integration}
\author{Victor Christen}
\orcid{0000-0001-7175-7359}
\affiliation{%
  \institution{Leipzig University \& ScaDS.AI}
  \streetaddress{Augustusplatz 10}
  \city{Leipzig}
  \country{Germany}
  \postcode{04109}
}
\email{christen@informatik.uni-leipzig.de}

\author{Peter Christen}
\orcid{0000-0003-3435-2015}
\affiliation{%
  \institution{Australian National University}
  \streetaddress{}
  \city{Canberra}
  \country{Australia}
  \postcode{}
}
\email{peter.christen@anu.edu.au}


\begin{abstract}
\Ac{er} is a fundamental task in data integration that enables insights from heterogeneous data sources. The primary challenge of ER lies in classifying record pairs as matches or non-matches, which in multi-source \ac{er} (MS-ER) scenarios can become complicated due to data source heterogeneity and scalability issues. Existing methods for MS-ER generally require labeled record pairs, and such methods fail to effectively reuse models across multiple \ac{er} tasks.
We propose \method (\textit{Model Repositories for Entity Resolution}), a novel method for building a model repository consisting of classification models that solve \ac{er} problems. By leveraging feature distribution analysis, \method clusters similar \ac{er} tasks, thereby enabling the effective initialization of a model repository with a moderate labeling effort. Experimental results on three multi-source datasets demonstrate that \method achieves comparable or better results to methods that have label-limited budgets, such as active learning and transfer learning approaches, while outperforming self-supervised approaches that utilize large pre-trained language models. When compared to supervised transformer-based methods, \method achieves comparable or better results, depending on the size of the training data set used. 
\end{abstract}

\maketitle

\section{Introduction}

\begin{figure}[t]
    \centering
    \includegraphics[width=\columnwidth]{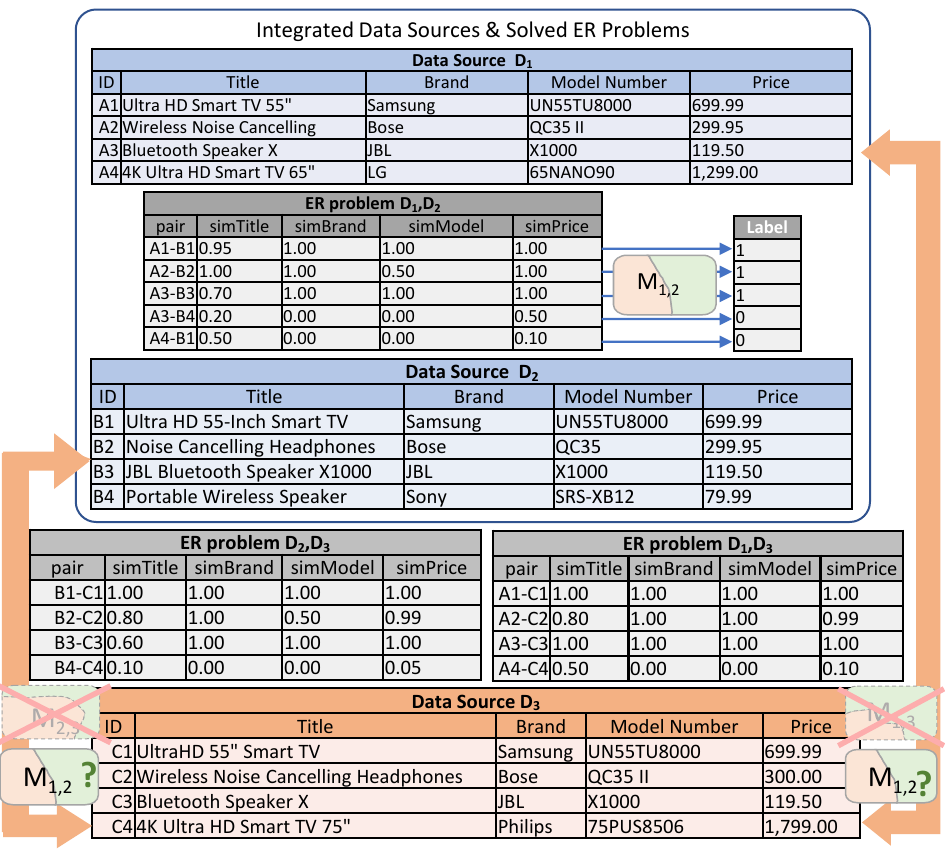}
    \caption{Motivation of reusing solved \ac{er} tasks for new tasks. The data sources $D_1$ and $D_2$ are already linked utilizing similarity feature vectors and a model $M_{1,2}$ to label each record pair. The question is whether the derived model $M_{1,2}$ can also be applied to the new data source, $D_3$, to match $D_3$ to $D_1$ and $D_2$, or if new models have to be generated.}
    \label{fig:motivation}
\end{figure}

Due to the increasing number of available data sources, data integration is essential for comprehensive analysis and a standardized view of information that is often scattered across sources~\cite{stonebraker2018di}. A cornerstone of data integration is \acf{er}, which identifies and links records that refer to the same real-world entity~\cite{Chr12}. The broad range of potential data quality issues and data source heterogeneity make \ac{er} challenging~\cite{Chr12, christophidesEP2021er_overview}. Multi-source \ac{er}~\cite{christophidesEP2021er_overview}, especially, amplifies multiple challenges that range from scalability issues to achieving high-quality linkage results.
 
The quality of the \ac{er} result between two data sources is highly dependent on the classification step~\cite{stonebraker2018di}. This step classifies record pairs into matches and non-matches based on attribute value similarities or derived embeddings from pre-trained language models. State-of-the-art methods use machine learning (ML) based approaches since they have been shown to achieve high-quality results~\cite{Doan20magellan}. However, the majority of ML methods require labeled training data to build a classification model, where generating such data can be a time-consuming and expensive task~\cite{Mudgal2018deepLearning}. 

Various methods have been proposed to reduce the labeling effort, such as \acf{al}~\cite{christen19infoal,Ngo12,Moz14,Primpeli21graphAl} and transfer learning~\cite{Kirielle2022TransER,Jin21DeepTrans, LosterKN21}. 
However, to the best of our knowledge, no existing method supports the reuse of trained \ac{er} classifiers by building and utilizing a model repository for \ac{er}.  


Typical use cases for model repositories are found in research institutions and companies that require the continuous integration of new and diverse data sources.
For example, in healthcare, hospitals and clinics often need to integrate patient records from various systems to create unified medical histories in order to enable better diagnosis, treatment, and research~\cite{Padmanabhan2019}. Similarly, in census data analysis, government agencies must reconcile records from multiple surveys and administrative sources to ensure accurate population statistics and policy planning~\cite{splink2022}. In e-commerce, product comparison portals and marketplaces integrate product catalogs from different vendors to provide search and recommendation functionalities for users.

In all these scenarios, our method will allow organizations to reuse previously trained models from solved \ac{er} tasks, significantly reducing both labeling efforts and computational costs. Furthermore, our method lays the groundwork for establishing \ac{er} matching services, enabling users to solve any \ac{er} problem by leveraging existing models. This capability is particularly valuable in dynamic environments where new data sources are continuously added, ensuring scalability and efficiency in real-world applications.

\Cref{fig:motivation} illustrates the motivation of our proposed method. Assuming we have two integrated data sources, $D_1$ and $D_2$, consisting of products specified by the shown attributes. The integration of these data sources requires the solution of the \ac{er} problem given by the classification task of the similarity feature vectors into matches and non-matches. These similarity features are derived from the attribute-wise comparisons between record pairs. Therefore, the classification model, $M_{1,2}$, based on labeled record pairs, is trained and used to classify the record pairs $A1-B1, A2-B2, A3-B3, A3-B4 \text{ and } A4-B1$. Later on, a new data source $D_3$ needs to be integrated, where we calculate similarity features between data sources $D_1$ and $D_3$, as well as between $D_2$ and $D_3$. 


In a naive solution, we would build one new classification model, $M_{1,3}$ and $M_{2,3}$, for each new \ac{er} task between the data sources $D_1$ and $D_3$ and $D_2$ and $D_3$, respectively. Due to the increasing number of models with the growing number of integrated data sources, the labeling effort for generating training data would, however, increase substantially. Instead of building the new models $M_{1,3}$ and $M_{2,3}$, our aim is to reuse the existing model $M_{1,2}$ that has already solved the (potentially similar) \ac{er} task between $D_1$ and $D_2$. As the amount of data to be integrated increases, it becomes crucial to efficiently generate training data for similar ER problems to ensure effective and scalable classification. Our proposed method, \method, addresses these challenges.

\begin{figure}[t]
    \subfloat[][Matches]{\includegraphics[scale=0.24]{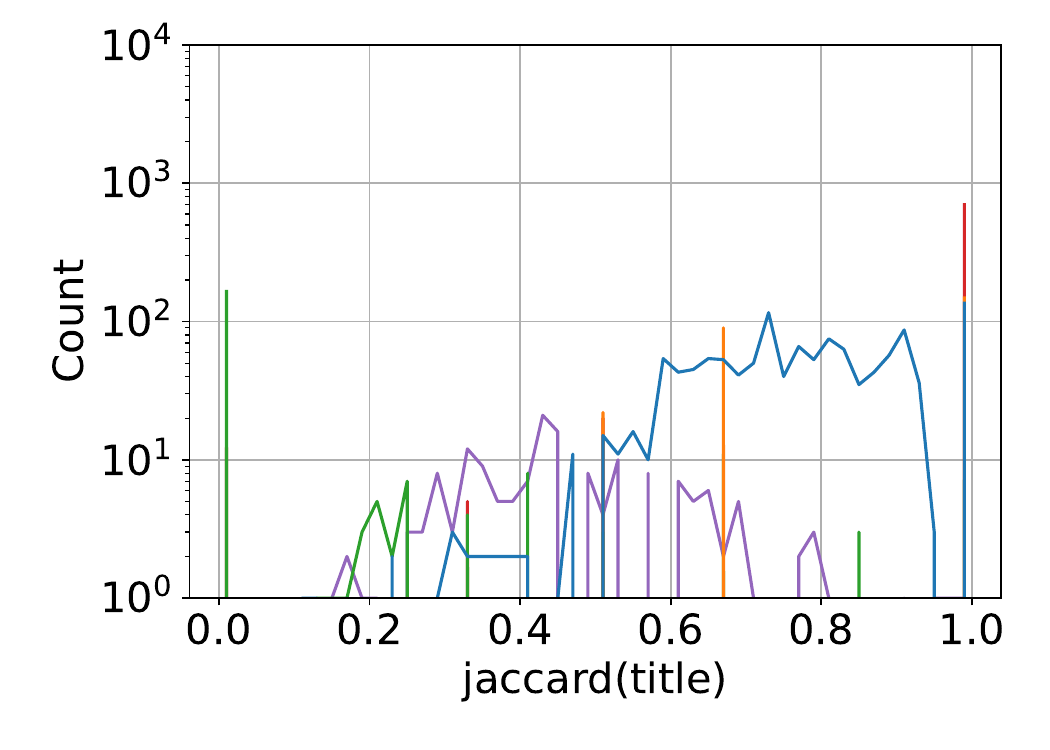}\label{fig:wdc_m}}
    \subfloat[][Non-Matches]{\includegraphics[scale=0.24]{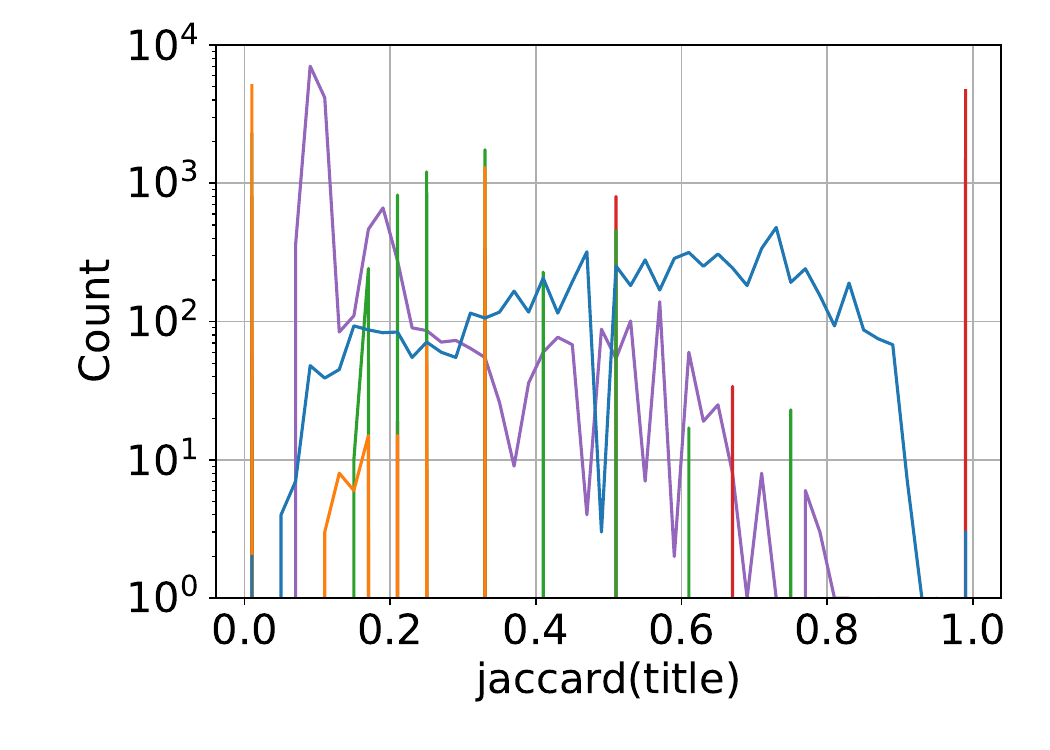}\label{fig:wdc_nm}}
    \caption{Example of the similarity distributions calculated using the Jaccard string comparison function~\cite{Chr12} for title considering the ER problems in the WDC computer data set. Each of the five lines represents an \ac{er} problem.}
    \label{fig:sim_distribution}
\end{figure}

An alternative approach would be to unify all existing \ac{er} problems to create a comprehensive training dataset. However, individual \ac{er} problems can have different distributions of their similarity features. This would make it challenging for a single classification model to account for the potentially diverse similarity distributions. To overcome this issue, the idea would be to create new training data for each new \ac{er} problem and combine these with existing one. However, such an approach is not scalable, since it requires the training data generation for each new ER problem.
\Cref{fig:sim_distribution} illustrates the variations in similarity distributions calculated by the Jaccard string comparison function~\cite{Chr12} on the title for products on the WDC computer dataset~\cite{Primpeli2019WDC}, which consists of four data sources. Given these very diverse similarity distributions, a single unified classification model would struggle to capture the diverse characteristics of matches and non-matches.
Moreover, new \ac{er} problems might not fit into an existing model, so retraining is required.

{To the best of our knowledge, 
no current approach takes model reuse into account. Existing approaches assume that training data are available to train a model and that this model can be used for all future record pairs, as in progressive or incremental
\ac{er}~\cite{Gazzarri2023edbt}.

Therefore, we suggest a method to create an \ac{er} model repository and to reuse the maintained models for similar future \ac{er} problems. The reuse of models is highly relevant where new data sources must be integrated in target data sources such as data warehouses or for ad hoc analysis with data from data lakehouses. 
}
In this paper, we make the following contributions:
\begin{itemize}
    \item We propose a novel method to build an \ac{er} model repository of classification models for \ac{er} problems. Our method analyzes the similarity feature space to determine the most appropriate model for an unsolved novel \ac{er} problem. {Unlike existing \ac{er} methods, our method distinguishes between \ac{er} tasks with different similarity feature spaces and selects the appropriate models for the various \ac{er} tasks.}
    \item Our proposed method efficiently initializes the repository with models that require only little labeling effort. To do so, we build an \ac{er} problem graph and cluster \ac{er} tasks according to their similar feature spaces. For each cluster, we build a classification model using the most relevant similarity feature vectors as training data. {The resulting clusters and models can be stored in a backend system and used for future \ac{er} tasks.}
    \item We evaluate our method using three multi-source \ac{er} datasets with up to 23 sources and compare it with state-of-the-art \ac{al} methods for multi-source \ac{er}, a transfer learning method, and pre-trained language model-based \ac{er} methods, considering the effectiveness and efficiency of all these methods.
\end{itemize}

The remainder of this paper is organized as follows. In Section~\ref{sec:Prelim} we formalize the problem we aim to solve, and in Section~\ref{sec:rel} we review related work on label-efficient and state-of-the-art ER methods.  In
Section~\ref{sec:reuseML} we then describe our method in detail, which we evaluate in Section~\ref{sec:eval} and compare it with several state-of-the-art approaches. In
Section~\ref{sec:discuss}, we discuss our results and provide recommendations. We conclude in
Section~\ref{sec:concl} with future research directions.


\section{Problem Definition}\label{sec:Prelim}

Our method aims to solve \ac{er} problems by using classification models from a repository. In real-world scenarios, continuous integration of new data sources is commonly required for ongoing analysis and maintenance~\cite{hai2023datalakes}.

Following the notation shown in~\Cref{tab:symbols}, we are given a set $\mathcal{D}$ of data sources $\{D_1,\ldots, D_z\}$. An \ac{er} problem $p_{k,l}$ is defined for a pair of data sources $D_k$ and $D_l$ consisting of similarity features for each record pair. We utilize an initial set $\mathcal{P_I}$ of \ac{er} problems $p_{i,l}$ to set up the \ac{er} model repository.
We aim to solve new \ac{er} problems $p_{i,l} \in \mathcal{P_U}$ utilizing the initial \ac{er} problems $\mathcal{P_I}$.

An \ac{er} problem $p_{i,l}$ is defined for a data source pair $D_i, D_l \in \mathcal{D}$ and consists of a set of similarity feature vectors $w$. 
 Each similarity feature vector $w \in p_{k,l}$ represents a record pair, where each element of $w$ corresponds to a similarity in the range of [0,1], calculated based on attribute value comparisons using a similarity function. We denote an element of a similarity feature vector $w$ as feature $f$. We assume that even for heterogeneous data sources, common features $f$ exist based on common attributes between the different data sources, which are significant to describe the entities of a certain domain, such as the title of products, price information, etc.
 In the example in~\Cref{fig:motivation}, $\mathcal{P_I}$ consists of the \ac{er} problem $P_{D_1, D_2}$ and $\mathcal{P_U}$  includes the \ac{er} problems $P_{D_2, D_3}$ and $P_{D_1, D_3}$. 

To reduce the effort of building new classification models and the underlying label generation process, our solution groups similar \ac{er} problems $p_{k,l} \in \mathcal{P_I}$. We determine the similarity between pairs of \ac{er} problems utilizing univariate distribution tests to group similar problems into clusters $\mathcal{C_{P}}$. We hypothesize that the \ac{er} tasks within a certain cluster $C\in \mathcal{C_{P}}$ can be effectively classified using a unified classification model $M_C$. 

The analysis uses the similarity distribution $d^f_{k,l}$ and the cumulative distribution function $CDF^f_{k,l}$ for each feature $f$ based on a data source pair $D_k$ and $D_l$. We construct an \ac{er} problem similarity graph $G_P=(\mathcal{P_I\cup \mathcal{P_U}}, E)$ where the edges represent the determined similarities between \ac{er} problems. The graph is used for clustering and is extendable to integrate new \ac{er} problems of $\mathcal{P_U}$.

\begin{definition} [Entity Resolution Model Search]
We define the problem of solving \ac{er} problems $p_{D_x, D_z} \in \mathcal{P_U}$ as the search for a similar cluster $C \in\mathcal{C_P}$ to apply the  model $M_C$ on the \ac{er} problem between $D_x$ and $D_z$. Moreover, we update the identified model $M_C$ if the \ac{er} problems of cluster $C$ become dissimilar or the coverage of training data of a certain cluster decreases below a certain threshold.
\end{definition}

\begin{table}[t]
    \centering
     \color{black}{
    \begin{tabularx}{\columnwidth}{lX}
    \toprule
       Symbol &Description \\
        $\mathcal{D}$ & Set of data sources $D_1,D_2,...D_z$\\
        $\mathcal{P_I}$ &Initial \ac{er} problems\\
        $\mathcal{P_U}$ &Unsolved \ac{er} problems\\
        $p_{i,k}$ & Set of similarity feature vectors $w$ between $D_i$ and $D_k$\\
        $G_P$ & \ac{er} problem similarity graph\\
        $\mathcal{C_P}$ & Set of clusters consisting of similar \ac{er} problems $p_{i,k}$\\
        $f$ & Feature representing a similarity function applied on corresponding attributes\\
        $w$ & Similarity feature vector consisting of similarities\\
        $d^f_{k,l}$ & Distribution of the feature $f$ from similarity feature vectors $w\in p_{k,l}$\\
        $CDF^f_{k,l}$& Cumulative distribution function for feature $f$ between data sources $D_k$ and $D_l$\\
          \bottomrule     
    \end{tabularx}
    }
    \caption{Description of used symbols.}
    \label{tab:symbols}
\end{table}

\section{Related Work}\label{sec:rel}

In this section, we discuss state-of-the-art \ac{er} approaches from traditional \ac{ml} based methods to language model based solutions such as utilizing LLMs. In general, the main goal of our method is to reduce the effort for the model generation step. Therefore, we discuss budget-limited \ac{ml} methods in \ac{er}, such as self-supervised, active, and transfer learning.

\Ac{er} methods classify record pairs based on features derived from similarity functions~\cite{Chr12}. In the past, supervised \ac{er} techniques have been developed, leveraging both traditional ML algorithms~\cite{Konda2016magellan, Kop10c, Bil03} and deep learning-based methods~\cite{Mudgal2018deepLearning, Ebraheem2017deepER}. The advent of transformer-based language models~\cite{Vaswani2017transformer} has further advanced \ac{er} by enabling the representation of textual and unstructured data as embeddings, which can be directly compared~\cite{LiDitto20, BrunnerS20}. Transformers, powered by self-attention mechanisms, process input data in parallel, making them highly effective for tasks like natural language processing and sequence modeling. State-of-the-art methods, such as \ditto~\cite{LiDitto20}, fine-tune transformer models like BERT~\cite{DevlinCLT19bert} and DistilBERT~\cite{Sanh2019distil} to classify record pairs as matches or non-matches. Auto-EM~\cite{AutoEM2019} relies on pre-trained models for different entity types to classify value pairs as match or non-match. For a new \ac{er} problem, the relevant entity types are identified, allowing the corresponding models to be utilized for classifying the record pairs. {\unicorn~\cite{Fan24Unicorn} generates an universal model addressing various matching tasks such as entity alignment, ontology matching, or \ac{er}, by using a unified encoder model and adopting a mixture-of-experts model.}
\multiem~\cite{zeng2024multiEM} clusters records from multiple data sources utilizing embeddings from pre-trained language models. This approach follows a hierarchical merge strategy where two data sources are compared to avoid the comparison of all data source pairs. While the proposed approach shows its strength in efficiency, it requires the identification of effective hyperparameters, the similarity threshold $m$, and clustering-related parameters. Peeters et al.~\cite{Peters2023GPT} showed the feasibility of using \acp{llm} for \ac{er} on various real-world datasets and different prompt variations without any fine-tuning. Recent work~\cite{ZhangAnyMatch2025} achieved comparable results by using problem-dependent fine-tuned small language models~\cite{ZhangAnyMatch2025} with a significant cost reduction. 



\smallskip
\textbf{Self-supervised Learning.}
Sudowoodo~\cite{Wang23Sudowoo} employs contrastive self-supervised learning to learn a similarity-aware data representation model for entities. The objective is to position similar entities closely together in the representation space while ensuring that dissimilar ones are placed further apart. 
The primary advantage of this approach is that it allows for the creation of meaningful data representations without the need for extensive labeled datasets. The fine-tuned model can be effectively utilized for various downstream tasks such as \ac{er}. ZeroER~\cite{Wu20zeroER} aims to learn the probability density functions of matches and non-matches based on the similarity features. The approach uses an adapted Gaussian Mixture model to determine the distributions of these similarity feature vectors for both classes (matches and non-matches). The adaptations address the overfitting issue and the transitivity constraint occurring in \ac{er}.

\smallskip
\textbf{Active Learning.}
\ac{al} approaches aim to minimize the labeling effort that is particularly important for multi-source \ac{er} due to the numerous classification tasks involved. Most \ac{al} methods focus on pairwise \ac{er} problems~\cite{Moz14, christen19infoal, Wan15, Ngo12} utilizing uncertainty measures and distances between feature vectors. Only one study~\cite{Primpeli21graphAl} (\emph{Almser}) has specifically addressed multi-source \ac{er} by leveraging similarity graphs generated from the \ac{er} process. This work employs the concept of transitive closure to identify informative links, enabling the detection of potential false negatives and false positives. Specifically, it assesses whether two records, classified as non-matches, are connected through the transitive closure. Potential false positives are identified using the minimum cut of the graph. At the same time, the resulting graph also facilitates the identification of false negatives, represented as missing edges among record pairs within connected components.

\smallskip
\textbf{Transfer Learning.}
Transfer learning methods aim to adapt a source domain to a target domain, minimizing the need for model training and data generation in the target domain. In the context of \ac{er}, these methods leverage existing models and training data from labeled \ac{er} tasks. For instance, Jin et al.~\cite{Jin21DeepTrans} developed a joint neural network with a shared attention mechanism to optimize the weighting of different attribute comparisons for both source and target \ac{er} tasks. Loster et al.~\cite{LosterKN21} proposed a method for transferring knowledge of a trained Siamese network to a target domain. The Siamese network learns record embeddings based on the given attributes utilizing a bidirectional LSTM
network (BLSTM)~\cite{SchusterP97blstm}. The architecture allows the transfer of embeddings and weights from the BLSTM layers encoding the attributes. 
Kirielle et al.~\cite{Kirielle2022TransER} et al. proposed a method for labeling record pairs in unsolved \ac{er} tasks, thereby generating a classification model suitable for traditional \ac{er} tasks. The authors identify similar feature vectors between source and target entity resolution tasks and utilize the labels from similar target vectors to construct the target model.

\smallskip
\textbf{Limitations}
Despite substantial progress in \ac{er} through machine learning, active learning, and transfer learning, existing approaches face several significant limitations, particularly in multi-source and dynamic integration settings. Most existing approaches focus on pairwise \ac{er}, requiring the generation of new classification models for each \ac{er} problem when the similarity feature space changes, resulting in a high training effort. In the following, we summarize the specific limitations for each area.

Most AL approaches are tailored to pairwise ER and do not generalize to the complexities of multi-source settings, where the number of possible classification tasks grows rapidly. Even recent multi-source AL methods, such as \almser \cite{Primpeli21graphAl}, focus on optimizing label selection within known tasks but lack mechanisms to determine when existing models can be reused or when new models (and thus fresh training data) are necessary for new \ac{er} problems. 

Transfer learning methods are limited by their inability to effectively select the most appropriate source task(s) when multiple candidates exist. Most transfer learning approaches require significant similarity between source and target domains, which is difficult to guarantee in heterogeneous and evolving multi-source environments. Furthermore, existing methods do not provide systematic procedures for evaluating model suitability.

While pretrained language model-based and \ac{llm} methods have substantially advanced ER performance, they struggle with practical constraints. Fine-tuning such models requires large labeled datasets, making them costly and impractical in scenarios with limited annotation resources. Moreover, LLM approaches are monetarily intensive and struggle to scale to the massive number of record pairs present in real-world multi-source ER tasks.  

No current method fully supports the efficient reuse of both models and labeled data across multiple, evolving ER tasks. In scenarios where data sources and their characteristics frequently change, existing approaches either require retraining from scratch or risk applying poorly matched models, leading to degraded performance. Methods that attempt to unify training data from multiple ER problems ignore the heterogeneity in feature distributions, which limits model generalizability and scalability.

\begin{figure*}[t!]
    \centering
    \includegraphics[width=0.9\textwidth]{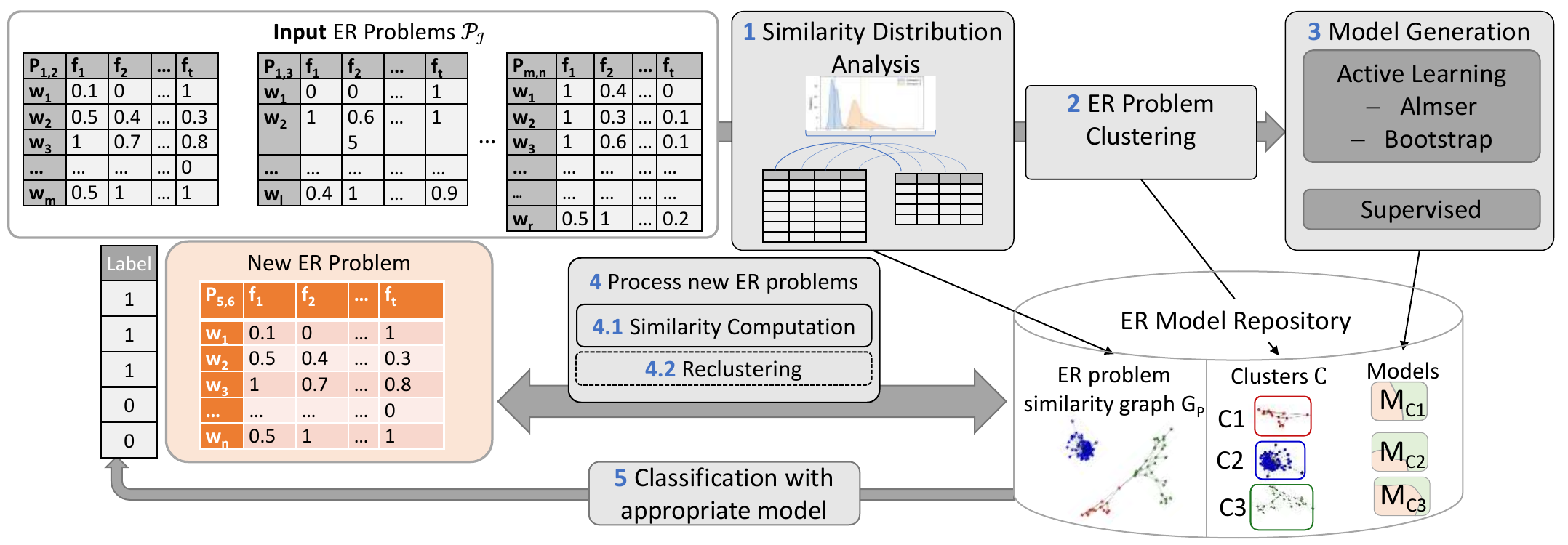}
    \caption{Workflow for initializing and using an \ac{er} model repository consists of the steps: 1. Similarity Distribution Analysis, 2. \ac{er} Problem Clustering, 3. Model Generation, 4. Process new \ac{er} problems, 5. Classification.} 
    \label{fig:workflow}
\end{figure*}
\section{Reusing of Solved ER Tasks}\label{sec:reuseML}

In this section, we introduce our novel method \method that takes advantage of previously solved \ac{er} problems, denoted as $\mathcal{P_I}$, along with their corresponding classification models. We initially give an overview and then describe the steps of \method in detail.

\subsection{Overview of \textit{MoRER}}
The input of our method is a set of \ac{er} problems that are used to initialise the repository. Our main focus is on reusing \ac{er} problems. As a reminder, we briefly describe how the similarity feature vectors $w\in p_{k,l}$ can be determined.
Theoretically, various blocking techniques~\cite{papadakisSTP20blocking} can be employed to reduce the search space based on the similarity calculations used. Recent approaches utilize pre-trained language models to generate embeddings representing records~\cite{LiDitto20, zeng2024multiEM}. In this case, nearest-neighbour search approaches~\cite{deepBlocker2021, Maciejewski2025proER} can be used to reduce the number of candidate record pairs~\cite{papadakisSTP20blocking}. 

The similarities can be computed using appropriate similarity functions tailored to the relevant attributes of the records being compared~\cite{Chr12}. A further strategy to leverage data sources with different attributes is the use of pre-trained language models. The models can generate embeddings that encompass all attributes, functioning similarly to a bag-of-words approach. Using this strategy, an \ac{er} can calculate embedding similarities for certain or all attributes.

Figure~\ref{fig:workflow} shows the workflow to efficiently create a model repository to solve new \ac{er} tasks of $\mathcal{P_I}$. We first conduct distribution analysis to determine whether the similarity distributions of features $f_1,\ldots,f_t$ from the \ac{er} problems are similar {considering univariate and multivariate analysis tests}. We calculate an similarity score, $sim_p$, for each $p_{k,l} \in \mathcal{P_{I}}$. 

From the pairwise comparison of \ac{er} problems, we derive the \ac{er} problem similarity graph $G_{P}$. The \ac{er} problems represent the vertices, and the similarities between the \ac{er} problems are the weighted edges. We cluster the graph $G_{P}$ using the Leiden algorithm~\cite{Traag2019leiden}. In pre-experiments, we also evaluated other algorithms such as Girvan-Newmann~\cite{Girvan_2002} and label propagation~\cite{RA07labelpropagation}, leading to similar results.
{As shown in \Cref{fig:workflow}, we calculate pairwise the similarities between $P_{1,2}, P_{1,3}\ldots P_{m,n}$ and compute clusters based based on the graph $G_P$ resulting in the clusters $C_1, C_2$, and $C_3$.}

We consider all \ac{er} tasks $p_{k,l}\in C^i$ as similar, assuming that one model is sufficient to classify the underlying similarity vectors from these \ac{er} tasks. Due to this simplification, we only need training data to generate one model for each cluster $C^i \in \mathcal{C_{P}}$.

The supervised solution in the model generation step considers all feature vectors of \ac{er} problems in $\mathcal{P_I}$ as training data, requiring high labeling effort. To reduce labeling effort, we apply active learning for each group to determine labeled training data~\cite{Primpeli21graphAl, Moz14}. The budget $b_{tot}$ constrains the total number of labeled data. Due to the skew of the size of clusters, a uniform distribution of the budget regarding the clusters is not useful. Therefore, we proportionally distribute the budget based on the number of similar feature vectors of the clusters. Moreover, we handle singleton clusters separately since we want to prioritize clusters with more than one \ac{er} problem. 

The result of the \ac{al} procedure for a cluster $C^i_{P}$ regarding the allocated budget is a labeled set of similarity feature vectors. We build a classification model with the obtained training data that we store for a certain cluster $C^i$. Moreover, we maintain the similarity feature vectors $P_{C^i}$ of the training data for each cluster to calculate similarities to new \ac{er} tasks. {Considering the example in \Cref{fig:workflow}, we generate the corresponding models $M_{C1}, M_{C2}$ and $M_{C3}$ for the three clusters. Each cluster is represented by the training data that has been used for the corresponding similarity calculations.}

Solving new \ac{er} problems $p_{x, z} \in \mathcal{P_U}$ requires selecting an appropriate model for each unsolved \ac{er} problem. We propose several strategies for this selection process. The first strategy, $sel_{\text{base}}$, identifies the most similar cluster $C^i$ and applies its associated model $M_{C^i}$. Specifically, the \ac{er} problem $p_{x, z}$ is compared with the set of similarity feature vectors \( P_{C^i} \) from each cluster \( C^i \in \mathcal{C_{P}} \). The cluster with the highest similarity score \( sim_p \) is selected, and its corresponding model \( M_{C^i} \) is used to classify the similarity vectors of \( p_{x, z} \) into matches and non-matches. {In the example, we calculate the similarities of the new \ac{er} problem $P_{5,6}$ to the cluster representatives $P_{C^1}, P_{C^2}$ and $P_{C^3}$.}

We address potential domain shifts that may arise when the unsolved \ac{er} problem is not adequately represented by existing clusters. To handle this, we propose the selection strategy $sel_{\text{cov}}$ described in~\Cref{sec:new_d} which integrates new \ac{er} problems and updates the existing models.

In the following section, we describe the method to compute similarities between \ac{er} problems. In detail, we provide a brief explanation of the distances used to compare two similarity distributions based on two features.  

\subsection{Similarity Distribution Analysis}\label{sec:stat_comp}
Initially, we compare pairwise the existing \ac{er} tasks $p_{k,l} \in \mathcal{P_I}$ of the data sources being initially integrated. {We consider univariate and multivariate statistical analysis.} For the univariate tests, we independently evaluate the distributions of each similarity feature $f_1,\ldots,f_t$. The application of univariate distribution tests is less complex and computationally expensive compared to multivariate tests. Moreover, we assume that the feature spaces are of the same size. However, for highly heterogeneous data sources, we recommend generating record embeddings based on the available attributes for each data source and calculating similarities between these embeddings covering multiple attributes.

{For the univariate tests, we} calculate a distribution similarity $sim_{dfi}$ for each feature pair and aggregate them into a weighted average similarity $sim_p$ using the standard deviation of each feature from the \ac{er} problems.

We consider the following distribution tests to determine if the similarities of a feature $f$ for the problems $p_{k,l}$ and $p_{m,n}$ are drawn from a similar distribution.

\smallskip
\textbf{Kolmogorow-Smirnow-Test.}
    This test~\cite{hollander2013ks} measures the maximum distance between the cumulative distribution functions $CDF^f_{k,l}$ and $CDF^f_{m,n}$ considering a certain feature $f$ of the \ac{er} tasks $p_{k,l}$ and $p_{m,n}$. The test statistic is defined by the supremum of the absolute difference between the two cumulative distribution functions given by~\Cref{eq:ks}.
    \begin{equation}
        KS=sup_{sim}|CDF^f_{k,l}(sim)- CDF^f_{m,n}(sim)|
        \label{eq:ks}
    \end{equation}
    
\textbf{Wasserstein distance.}
    This distance~\cite{villani2009ws} measures the cost to transform a distribution $Q$ into a distribution $W$. In our case, we consider the distributions $d^f_{k,l}$ and $d^f_{m,n}$ given by two column vectors for a feature $f$ of the entity resolution problems $p_{k,l}$, resp. $p_{m,n}$. Using these column vectors, the corresponding cumulative distribution functions \textcolor{black}{$CDF^f_{k,l}$ and $CDF^f_{m,n}$} are calculated and adapted to the same size if the column-wise feature vectors have different dimensions. The Wasserstein distance is the sum of the absolute differences between corresponding elements of $CDF^f_{k,l}$ and $CDF^f_{m,n}$, illustrated in~\Cref{eq:wd}. 
    \begin{equation}
        WD=\sum_{i=0}^{|CDF^f_{k,l}|} |CDF^f_{k,l}[i]-CDF^f_{m,n}[i]|\label{eq:wd}
    \end{equation}
    
\textbf{Population stability index.}
    This index~\cite{siddiqi2005psi} quantifies covariate shifts between two univariate distributions $d^f_{k,l}$ and $d^f_{m,n}$. Each distribution is split into bins $B_{k,l}$ and $B_{m,n}$ where the proportion $prop(b^i_{k,l})$ resp. $prop(b^i_{m,n})$ of similarities for the bins is determined. The index is defined as follows, where 100 is a commonly used number of bins $|B|$:
    \begin{equation} 
        PSI = \sum_{i=0}^{|B|} (prop(b^i_{k,l})-prop(b^i_{m,n}))\cdot ln\frac{prop(b^i_{k,l})}{prop(b^i_{m,n})} 
    \end{equation}
The resulting univariate distances of the features between two \ac{er} problems $p_{k,l}$ and $p_{m,n}$ are transformed into similarities and averaged as $sim_p$. Due to the different importance of similarity features, we weigh the average by utilizing the standard deviation of a similarity feature vector to consider the discriminative power of these features.

\smallskip
\textbf{Classifier two-sample test.} In addition to univariate distance measures, we employ a classifier two-sample test for multivariate distribution analysis~\cite{Lopez2016C2T}. The core idea is to train a classification model to distinguish between two datasets. If the distributions are similar, the classifier’s accuracy will be low, indicating that it struggles to differentiate between samples from the two datasets. We use the F1 score~\cite{Christen2023csur} to address the skewness according to the size of \ac{er} tasks. We define $sim_p$ as the inverse F1 score.

\subsection{Entity Resolution Task Clustering}
We utilize the determined aggregated similarity $sim_p$ between \ac{er} problems of $\mathcal{P_I}$ to build an \emph{entity resolution problem graph} $G_{P}=(\mathcal{P_I}, E)$.
Each \ac{er} problem represents a vertex, and an edge $e \in E$ represents the similarity between two \ac{er} problems. Each edge is weighted by the aggregated similarity $sim_p$. To prevent the formation of a single large connected component, we employ the Leiden algorithm~\cite{Traag2019leiden} to partition the graph into multiple clusters of similar \ac{er} problems, denoted as $\mathcal{C_{P}}$. 

Our method is flexible and can be adapted to incorporate other graph clustering methods~\cite{watteau2024graphclustering}, such as label propagation~\cite{Zhu2002labelprop} or the Girvan–Newman algorithm~\cite{Girvan_2002}. A key advantage of the Leiden algorithm is its ability to identify well-connected subgroups within weakly connected components~\cite{Traag2019leiden}. Furthermore, the algorithm is highly scalable~\cite{Traag2019leiden,watteau2024graphclustering}, making it suitable for large graphs, which is particularly relevant when dealing with a high number of \ac{er} problems.

\subsection{Model Generation}\label{subsec:model}
We consider \ac{al} approaches and supervised ML methods for generating the corresponding model for each cluster. For the \ac{al} approaches, we distribute a given budget $b_{tot}$ to the clusters.

\smallskip
\textbf{Budget distribution.} For each resulting cluster $C^i \in \mathcal{C_{P}}$, we generate a classification model $M_{C^i}$ to classify \ac{er} problems that are similar to those of cluster $C^i$. We limit the number of record pairs to build the training data by $b_{tot}$. We assign a minimum number of labels $b_{min}$ to each cluster to guarantee sufficient training data. 
To guarantee that the total budget is sufficient, we distinguish between singleton and non-singleton clusters $\mathcal{C_{P}}_{|s}$, respectively. $\mathcal{C_{P}}_{|ns}$. If~\Cref{eq:budget_req} holds, we have to merge the clusters of $\mathcal{C_{P}}_{|s}$ into clusters $C^i \in \mathcal{C_{P}}_{|ns}$ since the total budget is not sufficient for all clusters.
\begin{equation}\label{eq:budget_req}
    \mathcal{C_{P}} \cdot b_{min} > b_{tot}
\end{equation}
Otherwise, the remaining budget $b_{rem}$ determined by~\Cref{eq:rem_prop} is proportionally distributed to the clusters regarding the total number of non-singleton and singleton tasks. The ratio of the remaining budget for the non-singleton and singleton clusters is given by ~\Cref{eq:pp_1,eq:pp_2}. The budget for the singletons is distributed analogously.
\begin{equation}\label{eq:rem_prop}
    b_{rem}=b_{tot}-b_{min}\cdot|\mathcal{C_{P}}|
\end{equation}
\begin{equation}\label{eq:pp_1}
ratio(\mathcal{C_{P}}_{|ns}) = \nicefrac{|\mathcal{C_{P}}_{|ns}|}{|\mathcal{P}|}
\end{equation}
\begin{equation}\label{eq:pp_2}
ratio(\mathcal{C_{P}}_{|s}) = \nicefrac{|\mathcal{C_{P}}_{|s}|}{|\mathcal{P}|}
\end{equation}
The budget $b(C^i)$ for a cluster is calculated using ~\Cref{eq:dis_budget,eq:dis_budget2}, where $total_{C^i}$ represents the total number of feature vectors across all \ac{er} tasks within cluster $C^i$. If $C^i$ is a non-singleton cluster, its budget is the sum of $b_{min}$ and a proportion of the remaining budget $b_{rem}$ allocated among non-singleton clusters. The proportion for each non-singleton cluster is based on its number of feature vectors relative to the total number of feature vectors in all non-singleton clusters, as defined in the denominator in~\Cref{eq:dis_budget2}. In contrast, if $C^i$ is a singleton cluster, it receives a proportion of the remaining budget specifically allocated for singleton clusters, based on the total number of feature vectors from all singleton clusters.

\begin{equation}\label{eq:dis_budget}
    C^i_{|tot} = \biggl| \bigcup_{p_{k,l} \in C^i} p_{k,l}\biggr|
\end{equation}
\begin{equation}\label{eq:dis_budget2}
    b(C^i) = b_{min} + \frac{ C^i_{|tot}}{\sum_{C^k\in\mathcal{C_{P}}_{ns}} C^k_{|tot}}\cdot
     b_{rem} \cdot ratio(\mathcal{C_{P}}_{|ns}) 
\end{equation}

\textbf{Training data selection.}
A cluster $C^i$ consists of multiple \ac{er} problems that are similar to each other. We assume that one model can classify similar \ac{er} problems. Therefore, we train one classifier for each cluster. The model generation requires training data consisting of feature vectors labeled as matches and non-matches. In our method, we integrate two \ac{al} methods, \almser \cite{Primpeli21graphAl} (see~\Cref{sec:rel}) and the uncertainty \ac{al} method from Mozafari et al.~\cite{Moz14}. 
Note that other \ac{al} methods can be integrated. We have chosen \almser since it is the only method focusing on multi-source \ac{er}. Furthermore, the uncertainty method~\cite{Moz14} scales for large \ac{er} problems and simultaneously achieves qualitative results.

\smallskip
\textbf{Bootstrap.} The uncertainty method from Mozafari et al.~\cite{Moz14} utilizes a bootstrapping technique to compute an uncertainty measure. The process generates $k$ classifiers based on the current training dataset $\mathbf{T}$ by sampling with repetition. 
The determined models $m_1,...,m_k$  classify the unlabeled edge feature vectors, where the predictions are utilized to calculate the uncertainty $unc({w})$ of a similarity feature vector $w$ shown in \Cref{eq:base_unc}. The term $m_i({w})$ results in 0 or 1 if the feature vector $w$ represents a non-match or a match, respectively. 
\begin{equation} \label{eq:base_unc}
    unc({w}) = \frac{\sum_{i=1}^k m_i({w})}{k} \cdot \left(1-\frac{\sum_{i=1}^k m_i({w})}{k}\right)
\end{equation}

As an extension, we introduce a score $s$ for each similarity feature vector $w$ representing how unique a feature vector is regarding the associated records. The calculation of the score is similar to the inverse document frequency (IDF), considering the related records as words and the cluster as documents. We compute a record score $s_r$ and the derived score $s$ for a similarity vector $w$ as shown in~\Cref{eq:rec_score,eq:sf_score}. For each record, we determine the number of clusters where the record $r$ occurs $|\mathcal{C}_{\mathcal{P}|r}|$ and the total number of clusters. The score $s$ of a similarity feature vector $w$ is the average over the adjacent records $src$ and $tgt$ of the feature vector. 
\begin{equation}\label{eq:rec_score}
    \\
     s(w) = \nicefrac{[s_r(src(w)) + s_r(tgt(w))]}{2}
\end{equation}
\begin{equation}\label{eq:sf_score}
s_r(r) = log \frac{|\mathcal{C}_{\mathcal{P}|r}|}{|\mathcal{C_{P}}|}
\end{equation}

\smallskip
\textbf{Almser.} 
We use the original implementation\footnote{https://github.com/wbsg-uni-mannheim/ALMSER-GB} and extend it to support batch processing. As input, we only consider the induced subgraph regarding the \ac{er} tasks of a cluster. 

In addition to the model generation, we maintain the selected vectors by the \ac{al} method as the set $P_{C^i}$ for each cluster $C^i$. We utilize the sets to determine the best cluster for a new data source and the associated \ac{er} problems. 

\begin{figure}[t!]
   \includegraphics[width=0.9\columnwidth]{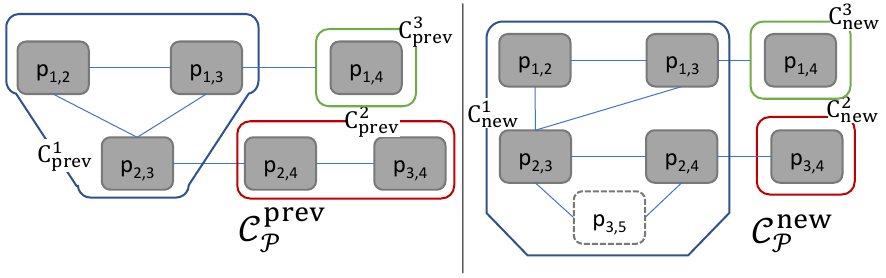}
    \caption{Example of integrating a new \ac{er} problem $p_{3,5}$. The grey colored \ac{er} problems represent problems of $T$.} \label{fig:recluster}
\end{figure}
\subsection{Solving New Entity Resolution Problems}\label{sec:new_d}
Integrating a new data source requires solving the \ac{er} problem $p_{x,z} \in \mathcal{P_U}$. We consider different strategies to determine an appropriate model. The first strategy, $sel_{\text{base}}$, assumes minimal domain shift, eliminating the need for new models or integration of additional \ac{er} problems. Here, the \ac{er} problem $p_{x,z}$ is compared with the set of feature vectors $P_{C^i}$ used to train models for each cluster $C^i$. We employ the same distribution test used to construct the \ac{er} problem graph $G_P$, incorporating the standard deviation of similarity vectors to weigh similarities. The cluster $C^i$ with the highest similarity $sim_p$ to $p_{x,z}$ is selected, and the corresponding model $M_{C^i}$ is applied to the feature vectors $w \in P_{x,z}$.

The second strategy $sel_{{cov}}$ addresses domain shifts in similarity feature distributions. For these methods, the new \ac{er} problem $p_{x,z} \in \mathcal{U}$ is compared with all \ac{er} problems in $\mathcal{P_I}$. We update the \ac{er} problem similarity graph $G_P$ by adding the \ac{er} problem $p_{x,z}$ and connecting it with the existing problems $p_{k,l} \in \mathcal{P_I}$. The graph is then reclustered using the Leiden algorithm already used before. We distinguish between two sets of \ac{er} problems: $T$, which includes problems already used for selecting training data, and $U$, which includes problems not yet used. A new model is trained if the unsolved problem $p_{x,z} \in C$ belongs to a cluster where all \ac{er} problems $p_{k,l} \in C$ are in $U$. Otherwise, the model $M_{C_{\text{prev}}}$ from the previous cluster $C_{\text{prev}} \in \mathcal{C}^{\text{prev}}_\mathcal{P}$ is reused. Reclustering may result in new clusters $C^{\text{new}}$ where \ac{er} problems from different previous clusters are grouped. For example, as shown in~\Cref{fig:recluster}, $p_{2,4} \in C^2_{prev}$ from the previous cluster is reassigned to $C^1_{new}$ in the new clusters $\mathcal{C}^{\text{new}}_\mathcal{P}$. In such cases, the model $M_{C_{\text{prev}}}$ from the cluster $C_{\text{prev}}$ with the maximum overlap according to $C_{\text{new}}$ is selected. 

As the number of new similarity vectors increases, it becomes more likely that the associated model $M_C$ for a cluster $C$ will potentially lead to the underrepresentation of the new \ac{er} problems within the cluster. Therefore, we define a coverage ratio $cov$ where we update the model $M_{C_{\text{prev}}}$ if $cov$ exceeds a predefined threshold $t_{\text{cov}}$. We compute the coverage ratio as the proportion of similarity feature vectors from \ac{er} problems in $U$ relative to all similarity feature vectors in cluster $C^{\text{new}}$.
    \begin{equation}
        cov(C) = \frac{\{w \mid w \in U \cap C\}}{\{w \mid w \in C\}}
    \end{equation}
\textcolor{black}{
Updating a model requires new training data. Therefore, we use similarity feature vectors from \ac{er} problems $p_{k,l} \in U \cap C^{\text{new}}$. We use the same \ac{al} method as for $\mathcal{P_I}$ (see \Cref{subsec:model}) with the following total budget to select the training data:
\begin{equation}
    b_{\text{new}} = b_{\text{tot}} \cdot cov(C) \cdot \nicefrac{|\{w \mid w \in T \cap C^{\text{prev}}\}|}{b_{\text{tot}}}
\end{equation}
We derive the budget from the originally used one $b_{tot}$, the coverage ratio $cov(C)$ defined above, and the ratio of record pairs used to train the previous model $M_{C_{\text{prev}}}$ to the total training budget. After updating the model, the set of \ac{er} problems used for training $T$ is extended and is removed from $U$. The updated graph $G_P$ and new clusters $\mathcal{C}^{\text{new}}_\mathcal{P}$ are then used for new problems of $\mathcal{P_U}$.
}

\section{Evaluation}\label{sec:eval}
In this section, we evaluate our method \method using three datasets, considering the multi-source \ac{al} method \almser\cite{Primpeli21graphAl}, the transfer learning method \transer\cite{Kirielle2022TransER} focusing on relational \ac{er} tasks, the self-supervised approach \sudowoodo\cite{Wang23Sudowoo}, the SOTA supervised \ac{er} approach \ditto\cite{LiDitto20}, and the small language model-based method \any\cite{ZhangAnyMatch2025}. Moreover, we analyze the components of \method such as the various distribution tests 
and the impact of retraining.

\subsection{Datasets}
Our method focuses on multi-source \ac{er} approaches, where the availability of datasets is limited. We consider three datasets being widely used in this domain~\cite{saaedi2018famer, zeng2024multiEM, Primpeli21graphAl, Pardo2025Gral, LermSR2021} as summarized in~\Cref{tab:stats}.

{\textit{Dexter.}} This dataset is derived from the camera dataset of the ACM SIGMOD 2020 Programming Contest. The dataset consists of 23 sources with around 21,000 records and intra-source duplicates. Each data source consists of source-specific attributes. 

{\textit{WDC-computer.}} We utilize the WDC training corpus for large-scale product matching and use the same subset derived from the original one being used in the \almser study~\cite{Primpeli21graphAl}, consisting of four data sources. 

{\textit{Music.}}
The MusicBrainz dataset is a synthetically generated dataset from the MusicBrainz (https://musicbrainz.org/) database. The dataset was corrupted by~\cite{Hildebrandt2020} and it consists of five sources with duplicates for 50\% of the original records. Each data source is duplicate-free, but the records are heterogeneous regarding the characteristics of attribute values, such as the number of missing values, the length of values, and the ratio of errors.

\subsection{Experimental Setup}
We implemented \method in Python 3.12 using scikit-learn 1.5.1 and networkx 3.3. We modified the existing \almser implementation as available on GitHub and reimplemented the uncertainty active learning method from Mozafari et. al.~\cite{Moz14}, which we called \textit{Bootstrap}. We set \textit{k}=100 achieving qualitative results following previous studies~\cite{christen19infoal,Moz14}. {We evaluate the effectiveness by calculating precision, recall, and F1 score~\cite{Christen2023csur} according to the predicted matches across overall \ac{er} tasks for a test dataset.}

All experiments were performed on a computer with a 64-core AMD(R) EPYC(R) 7713 2.0GHz - Turbo 3.7GHz, 50GB  allocated RAM, and an Nvidia Tesla A30. In the following, we will describe the generation of similarity feature vectors and the considered parameters of \method.

\smallskip
\textbf{Similarity Feature Vectors.} The main focus of our method is the classification and, consequently, the model generation and selection. Therefore, we pre-calculate the similarity vectors of the \ac{er} problems. We use the same attribute comparisons, {string similarity functions, }and blocking keys for the \dexter dataset as in previous studies~\cite{LermSR2021, SaeediDR2021}.
For the \wdc and \music datasets, we utilize the feature vectors\footnote{\url{http://data.dws.informatik.uni-mannheim.de/benchmarkmatchingtasks/almser_gen_data/}} used in the previous work presenting \almser~\cite{Primpeli21graphAl}. The authors use various string similarity functions for textual data and normalized  differences for numerical values.

\begin{table}[t]
 \caption{Statistics of the datasets used in our evaluation.}
    \label{tab:stats}
    \centering
    \begin{tabular}{lccc}
    \toprule
Name & \# ER problems & \# Record pairs & \# Matches \\
\midrule
\underline{D}exter & 276 & 1,100K & 368K \\
\underline{W}DC-computer & {12} & {74.5K} & {4.8K} \\
\underline{M}usic &  {20} & {385.9K} & {16.2K}  \\
\bottomrule
\end{tabular}
\end{table}

\smallskip
\textbf{Method Parameters.} We use the following configuration summarized in~\Cref{tab:parameters}, where the values written in bold are default values. 
\begin{table}[t]
\caption{Overview of the parameter setting for MoRER. The values in bold represent the default values.}\label{tab:parameters}
\centering
\begin{tabular}{cc}
\toprule
\textbf{Name}& \textbf{Values}\\ 
\midrule
$ratio_{init}$ & \textbf{50\%}, 30\%\\
Distribution test & \textbf{KS}, WD, PSI, C2ST\\
Model generation & \textbf{\ac{al}}, supervised\\
\ac{al} method & Bootstrap~\cite{Moz14}, Almser~\cite{Primpeli21graphAl}\\
Selection method & $\mathbf{sel_{base}}$, $sel_{cov}$\\
\bottomrule

\end{tabular}

\end{table}
We split the available \ac{er} problems into 50\% initial tasks $\mathcal{P_I}$ and consequently 50\% unsolved \ac{er} tasks $\mathcal{P_U}$ for the \dexter dataset. Note that we also consider the linkage between the same data sources since the data sources are not duplicate-free for the \dexter dataset. For the \wdc and \music datasets, we utilize the train-test split of the provided feature vectors. If a data source pair $(D_1, D_2)$ consists of record pairs $(r_k, r_l)\in train$ and $(r_m, r_n) \in test$ being present in the training and test dataset, we built two \ac{er} problems with data source pairs $(D_{1train}, D_{2train})$ and $(D_{1test}, D_{2test})$. 

\begin{table*}[t]
\small
\setlength{\tabcolsep}{1.3pt}
\centering
\caption{Linkage quality (Precision/Recall/F1 score) comparison of the budget limited methods \textit{MoRER} in combination with an \ac{al} method to \textit{Almser} standalone, \textit{Sudowoodo}, and \textit{AnyMatch}, and the supervised methods \textit{MoRER}, \textit{TransER}, and \textit{Ditto}. We highlight the best F1 score in bold and underline the second best one for each budget and method type (\ac{al} vs supervised). }\label{tab:comp}
\begin{tabular}{ll|c|c|c|c|c||l|c|c|c|c}
{D} & B & MoRER+Almser & MoRER+BS & Almser & Sudowoodo  & AnyMatch & B & MoRER & Ditto & Unicorn& TransER\\
\hline
\multirow{3}{*}{D}    &1K  & 0.96/0.92/\textbf{0.94}      & 0.88/0.90/0.89 & 0.97/0.87/\underline{0.92} & 0.42/0.76/0.54 & 0.59/0.96/0.73 &    50\%&0.94/0.92/\textbf{0.93} & 0.60/0.99/0.75 & 0.53/0.96/0.68 &0.95/0.90/\underline{0.92}\\
                        &1.5K& 0.96/0.92/\textbf{0.94}    & 0.85/0.94/0.89 & 0.97/0.89/\underline{0.92} & 0.43/0.68/0.53 & 0.60/0.97/0.74 &    all &  0.91/0.93/\textbf{0.92} & 0.60/0.99/0.75  & 0.56/0.93/0.70 & 0.94/0.87/\underline{0.90}\\
                        &2K  & 0.96/0.90/\underline{0.93} & 0.88/0.93/0.90 & 0.97/0.90/\textbf{0.94}    & 0.45/0.85/0.58 & 0.59/0.98/0.73 &             &  &  & \\
\hline
\multirow{3}{*}{W}  &1K   & 0.84/0.83/\underline{0.83}   & 0.85/0.77/0.81 & 0.93/0.92/\textbf{0.93}  & 0.43/0.99/0.60  & 0.55/0.97/0.70     &   50\%&0.94/0.93/\textbf{0.94} & 0.92/0.95/\underline{0.93} & 0.77/0.92/0.84 & 0.79/0.86/0.82 \\
                    &1.5K & 0.94/0.87/\underline{0.90} & 0.90/0.86/0.88 & 0.94/0.93/\textbf{0.93} & 0.40/0.99/0.57  & 0.58/0.94/0.72   &       all &0.94/0.92/\underline{0.93} & 0.98/0.94/\textbf{0.96}& 0.90/0.90/0.90 & 0.80/0.86/0.83\\
                    &2K   & 0.95/0.94/\textbf{0.94} & 0.79/0.84/0.84 & 0.95/0.93/\textbf{0.94}  & 0.47/0.90/0.61  & 0.61/0.98/0.75   &       & &  & \\
\hline
\multirow{3}{*}{M}
    &1K  & 0.98/0.96/\underline{0.97}  & 0.94/0.92/\underline{0.97} & 0.98/0.97/\textbf{0.98} & 0.92/0.95/0.94  & 0.97/0.99/\textbf{0.98} &   50\% & 0.99/0.96/{0.97} & 0.99/0.99/\textbf{0.99} & 0.98/0.98/\underline{0.98} & 0.96/0.93/0.94\\
    &1.5K& 0.98/0.97/\textbf{0.98}& 0.98/0.97/\underline{0.97} & 0.99/0.97/\textbf{0.98} & 0.95/0.96/0.96  & 0.93/0.99/0.96 &   all & 0.99/0.97/\underline{0.98} & 1.00/0.99/\textbf{0.99} & 0.97/0.99/\underline{0.98}& 0.95/0.92/0.94\\
    &2K  & 0.99/0.96/\textbf{0.98}& 0.96/0.96/\underline{0.97} & 0.99/0.97/\textbf{0.98} & 0.93/0.96/0.95  & 0.96/0.99/\underline{0.97} &   &      &  & \\
\bottomrule
\end{tabular}

\end{table*}
\begin{figure*}[t!]
    \centering{\includegraphics[width=0.96\textwidth]{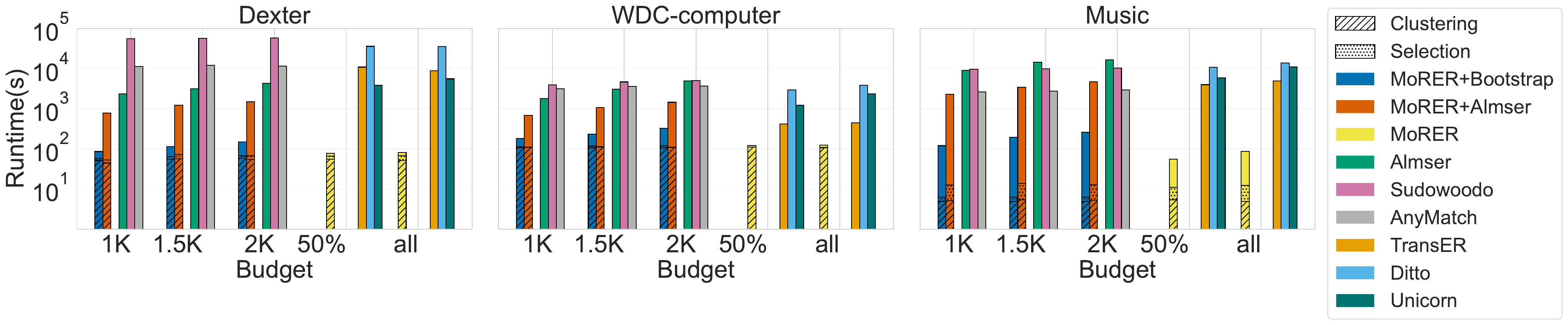}}
    \caption{Runtime comparison of \textit{MoRER} with \ac{al} and the supervised version to \textit{Almser}, \textit{Sudowoodo}, \textit{AnyMatch}, \textit{TransER}, and \textit{Ditto}. The shaded areas represent the runtime for the statistical analysis/clustering (striped) and the selection (dotted) of a model.}
    \label{fig:runtime_comparison}
\end{figure*}

\smallskip
\textbf{Compared methods.}
Due to the limitations of existing methods for reusing ML models, we focused on label-efficient techniques like active learning and transfer learning. Our method is compared to \almser~\cite{Primpeli21graphAl}, and \transer~\cite{Kirielle2022TransER}. All methods utilized identical similarity feature vectors, with a consistent division between solved and unsolved \ac{er} tasks to ensure comparability. 

We also compare our method against language model-based approaches, which have shown superior performance over traditional methods for processing textual data. To generate record embeddings, we utilize the same set of attributes originally employed to create the similarity vectors. 
We consider 
\sudowoodo~\cite{Wang23Sudowoo}, \ditto~\cite{LiDitto20}, {\unicorn~\cite{Fan24Unicorn}}, and \any~\cite{ZhangAnyMatch2025}.

We utilize the same record pairs for the training and test sets for each method. We partition the training and test data sets by the source pairs of the corresponding record pairs to generate \ac{er} problems for $\mathcal{P_I}$ and $\mathcal{P_U}$.
Unlike \ac{al} methods (such as \emph{Almser}), \transer, \ditto, {and \unicorn} are not designed to select informative links. Instead, we treat all initial \ac{er} tasks $\mathcal{P_I}$ among the integrated data sources as training data and 50\% of the similarity feature vectors for training. This consideration results in the following training data sizes of all pairs for the \dexter, \wdc, and \music datasets: 68,339, 231,521, and 46,049 record pairs, respectively. These training sizes reflect the volume of labeled record pairs used by \transer, \ditto, {and \unicorn} for their classification tasks. For \sudowoodo and \any, we use the semi-supervised approach with the same labeling budget as our method \method and \almser.

\almser\cite{Primpeli21graphAl}. We utilize the integrated data sources and their corresponding \ac{er} tasks, maintaining the same number of labeled pairs as our method. To ensure fairness, we modified the original implementation\footnote{\url{https://github.com/wbsg-uni-mannheim/ALMSER-GB}} to support batch processing and use graph-inferred labels from cleaned connected components as per the original study.

\transer\cite{Kirielle2022TransER}. We use the implementation available on GitHub\footnote{https://github.com/nishadi/TransER}. \transer's parameter values are set at~\cite{Kirielle2022TransER} $k=10$, $t_c=0.9$, $t_l=0.9$, and $t_c=0.9$, as determined through pre-experiments. In this context, $k$ represents the neighborhood size of a feature vector, while $t_c$ and $t_l$ are the thresholds for class confidence similarity and structural similarity between the neighborhoods of the source and target domains. Additionally, $t_p$ denotes the threshold for the confidence of pseudo labels.

\sudowoodo\cite{Wang23Sudowoo}. We use the available implementation on GitHub\footnote{https://github.com/megagonlabs/sudowoodo} with the default parameters presented in the original work~\cite{Wang23Sudowoo}. Moreover, we consider all proposed optimizations showing the best results according to the different datasets. For comparability, we use the semi-supervised variant with the same labeling budget. 

\ditto\cite{LiDitto20} {and \unicorn\cite{Fan24Unicorn}}. For faster training, we utilize DistilBERT~\cite{Sanh2019distil} as a pre-trained model. Moreover, for \ditto, we use a maximum sequence length of 256, also fixed in the original work~\cite{LiDitto20}, and train the model for 10 epochs. We report the results that lead to the smallest training loss over the 10 epochs. {For \unicorn, we use the default parameter values of the implementation on GitHub\footnote{\url{https://github.com/ruc-datalab/Unicorn}} with six experts. We utilize the combined loss function covering the diversity and a balanced importance of experts.}

We evaluate \any\cite{ZhangAnyMatch2025} using its official GitHub implementation\footnote{\url{https://github.com/Jantory/anymatch}}. For each dataset, the model is trained on the corresponding training set and evaluated on the test set, with the parameterized sample size $n_r$ for comparability. Note, \any filters relevant record pairs using the full ground truth of the training data. We use a batch size of 32, GPT-2 as the language model, and a learning rate of $2^{-5}$. 

\subsection{Comparison to Baselines}

\smallskip
\textbf{Effectiveness.}
The results, shown in~\Cref{tab:comp}, indicate that \method performs comparably to \almser under certain budgets and active learning configurations. 
\method, when combined with \almser, outperformed both the \textit{Bootstrap} method and \almser alone on the \dexter dataset. We hypothesize that \almser benefits from the high number of entity resolution tasks, resulting in rich similarity graphs, even when these graphs are divided due to the separation of \ac{er} tasks. Moreover, the partitioning of \ac{er} tasks reduces the search space for identifying informative links for similar classes of entity resolution tasks, thereby minimizing the likelihood of selecting uninformative feature vectors.

\almser outperforms \method+\textit{Bootstrap} by up to 9\% for budgets of 1000 and 1500. For the \dexter dataset, \method's average F1 score using the Bootstrapping-based AL method was approximately 4.3\% lower than \almser's. For the \music dataset, all \method combinations and \almser achieve similar results, with F1 score differences of at most 0.01 for a budget above 1500.  

In contrast, \transer consistently underperformed relative to the combined performance of \almser and \method across all tasks, regardless of whether the full training dataset or the 50\% version was utilized. This consistent underperformance shows its limitations in handling large volumes of heterogeneous training data, which is particularly evident in the \dexter and \wdc datasets. 
Considering the language model-based methods without using labeled training data, \sudowoodo and \multiem achieve comparable results compared to the aforementioned methods according to the \music dataset. However, the approaches result in a decreased F1 score of up to 38\% (\textit{Dexter}) when comparing \sudowoodo with \method+\almser.
\sudowoodo was originally conceptualized for pairwise record linkage. Therefore, we assume that the various and 
heterogeneous \ac{er} problems in \dexter and \wdc cause the challenge of determining representations to distinguish matches from non-matches.

\begin{table*}[t!]

\setlength{\tabcolsep}{1.5pt}
 \small
  \caption{Summary of speedup factors for \textit{MoRER} with various AL methods compared to the approaches \textit{(Alm)ser}, \textit{TransER} (TER), \textit{(Su)dowoodo}, \textit{(Dit)to}, \textit{(Uni)corn}, and \textit{(Any)Match} regarding the datasets \textit{(Dex)ter}, \textit{(Mus)ic}, and \textit{(WDC)-computer}.}
    \label{tab:rounded_budget}
    \centering
\begin{tabularx}{\textwidth}{lrrrrrrrrrrrrrrrrrrrrrrrrrrr}
\toprule
        &  \multicolumn{9}{c}{Budget: 1000} & \multicolumn{9}{c}{Budget: 1500} & \multicolumn{9}{c}{Budget: 2000} \\
\cmidrule(lr){2-10} \cmidrule(lr){11-19} \cmidrule(lr){20-28}
DS  & Alm & TER & TER &Su & Dit & Dit &Uni & Uni & Any& Alm & TER & TER & Su & Dit & Dit &Uni & Uni& Any& Alm & TER & TER &Su & Dit & Dit&Uni & Uni & Any \\
&  &  50\% & all & &  50\%&all &  50\% & all & & &  50\% & all &  &  50\% &all &  50\% & all & & &  50\%  & all & &  50\% &all &  50\%& all & \\
\midrule
\multicolumn{28}{c}{\method+\almser}\\
Dex   & 3.0    & 14.1  & 11.3   & 72  & 45.7   & 45.3  & 5  & 7.2 & 14.6  & 2.6   &8.9     & 7.2  & 46   & 29.0 & 28.8  & 3.2  & 4.6 & 9.8   & 2.9    & 7.3    & 5.9  & 38.3  & 23.6   & 23.4& 2.6  & 3.7 & 7.8    \\
Mus   & 4.0    & 1.7   & 2.1    &4.2  & 4.6    & 6.0   & 2.6   & 4.9 & 1.1   & 4.3   & 1.2    & 1.4  & 2.9  & 3.1  & 4.1   & 1.7   & 3.3 & 0.8   & 3.5    & 0.8    & 1.0  & 2.2   & 2.2    & 2.9 & 1.2   & 2.3 & 0.6    \\
WDC   & 2.6    & 0.6   & 0.6    &5.8  & 4.3    & 5.6   & 1.8   & 3.4 & 4.6   & 2.9   & 0.4    & 0.4  & 4.4  & 2.7  & 3.6   & 1.2   & 2.2 & 3.4   & 3.4    & 0.3    & 0.3  & 3.5   & 2.0    & 2.6 & 0.9   & 1.6 &2.5   \\
\midrule
\multicolumn{28}{c}{\method+\textit{Bootstrap}}\\
Dex  & 27.3 & 127.3 & 102.7 & 46   & 413.8 & 410.5 & 45.1  & 65 & 132.2& 28.1 & 96.6 & 78.0 & 41.2  & 314.1 & 311.6& 34.3  & 49.4 & 106.1 & 28.9  & 72.8 & 58.7  & 33.4  & 236.5 & 234.6 & 25.8 & 37.2  & 76.8   \\
Mus  & 75.7 & 33.1  & 40.6  & 80.6 & 87.4  & 115.4 & 49   & 92.8 & 74.3 & 21.9 & 20.2 & 24.8 & 49.9  & 53.5  & 70.6 & 30 & 56.7 & 14    & 63.8  & 15.3 & 18.8  & 40    & 40.4  & 53.4  & 22.7  & 42.9 & 11.2  \\
WDC  & 9.8  & 2.3   & 2.4   & 306  & 16.0  & 21.0  & 6.8   & 12.8 & 17.3 & 13.2 & 1.8  & 1.9  & 240.5 & 12.5  & 16.4 & 5.3   & 10 & 15.4  & 15.0  & 1.3  & 1.4   & 176.7 & 8.9   & 11.8  & 3.8  & 7.2  & 11.2  \\
\bottomrule
\end{tabularx}
\end{table*}

\any and \unicorn achieve comparable performance only on the \music dataset. For the \wdc dataset, the results of \any decrease by up to 22\% compared to \method+\almser and by 17.4\% when using \method with \textit{Bootstrap}. We assume this decrease is caused by the limitations of the AutoML selection, which appears insufficient for building effective models when classifying large numbers of record pairs. {For \dexter and \wdc, \unicorn generates results that decreased by up to 25\% in F1 score compared to \method+supervised.}

 \ditto, when utilizing the complete training datasets, outperforms the \method+\almser method by an average of 3\% and the \textit{Bootstrap} combination by up to 6\% for budgets exceeding 1500. However, for the \dexter dataset, the \method combinations achieved higher F1 scores than \ditto. The \dexter dataset, comprising camera specifications with numerous model numbers, introduces challenges due to minor textual differences that can lead to non-matches. We hypothesize that pretrained language model-based methods, such as \ditto, struggle with accurately classifying matches and non-matches when only small textual distinctions exist, particularly in scenarios with limited training data. We can also observe the importance of the training data size, considering the results achieved with 50\% of the training data. The difference in F1 score is reduced, or \method achieves higher F1 scores, as observed in the \wdc dataset with a budget of 2000.
 
\smallskip
\textbf{Efficiency.} {We consider the total runtime that encompasses training data generation, model training, and classification. Notably, for AL methods, the time required to select training data significantly impacts the overall labeling process, since the selection of informative samples often is complex and computationally intensive.} \method combined with the \textit{Bootstrap} \ac{al} method demonstrates significant advantages, outperforming both \almser and \transer across all datasets and labeling budgets, as shown in ~\Cref{fig:runtime_comparison}. {The additional effort required for statistical analysis, \ac{er} clustering and model selection is minimal, ranging from 6 seconds (\music) to 121 seconds (\wdc). Compared to selecting training data using an \ac{al} method, the proportion ranges from 0.3\% (\music with \almser) to 67\% (\textit{Dexter} with Bootstrap).} 

For example, on the \music dataset, \method+\textit{Bootstrap} runtimes range from 121 to 2621 seconds, whereas \almser can take up to 4.6 hours. \almser's longer runtimes are due to the increasing size of similarity graphs, which require more computational resources for graph analysis. The efficiency advantage results from the \method strategy of clustering similar \ac{er} tasks among the integrated data sources to select training data. This clustering approach significantly decreases the search space for potential links, contributing to the observed performance improvements. Reducing the search space also positively influences the combination of \method and \almser. The performance boost results in a maximum speedup of 4.3x (\music dataset, budget = 1000) over \almser alone. \transer's runtimes range from approximately 7 minutes (\wdc) to 2.5 hours (\dexter), considering the complete training dataset and 50\% of it.

For datasets consisting of many feature vectors, the clustering approach in \method reduces the search space for selecting training data. In contrast to \method, \transer compares each unsolved feature vector with all feature vectors from the integrated \ac{er} tasks. 
 \ditto is slower than \method combined with \almser or \textit{Bootstrap} due to its  training times. In contrast, \any and \unicorn achieve smaller runtimes by using smaller training datasets {resp. less complex models}. However, neural network-based methods, including \ditto, \unicorn, and \any, generally require more training time than non-language model-based approaches such as \method, \almser, and \transer. Notably, \sudowoodo exhibits the longest runtimes, reaching up to 16 hours on the Dexter dataset.


\Cref{tab:rounded_budget} presents a summary of the speedup factors for \method in comparison to other supervised \ac{ml} methods.
The analysis highlights that using \textit{Bootstrap} as an \ac{al} method within \method results in significant runtime improvements over \almser, especially when compared to \almser alone, \transer, \ditto, and \sudowoodo. In terms of effectiveness, \method combined with the \textit{Bootstrap} \ac{al} method leads to a decrease in quality compared to the \almser combination. Due to the missing training data generation step, the supervised-based \method achieves the best speedup compared to the other approaches.

\begin{figure*}[t!]
    \centering
    \begin{subfigure}[t]{5.6cm}
        \includegraphics[width=\linewidth]{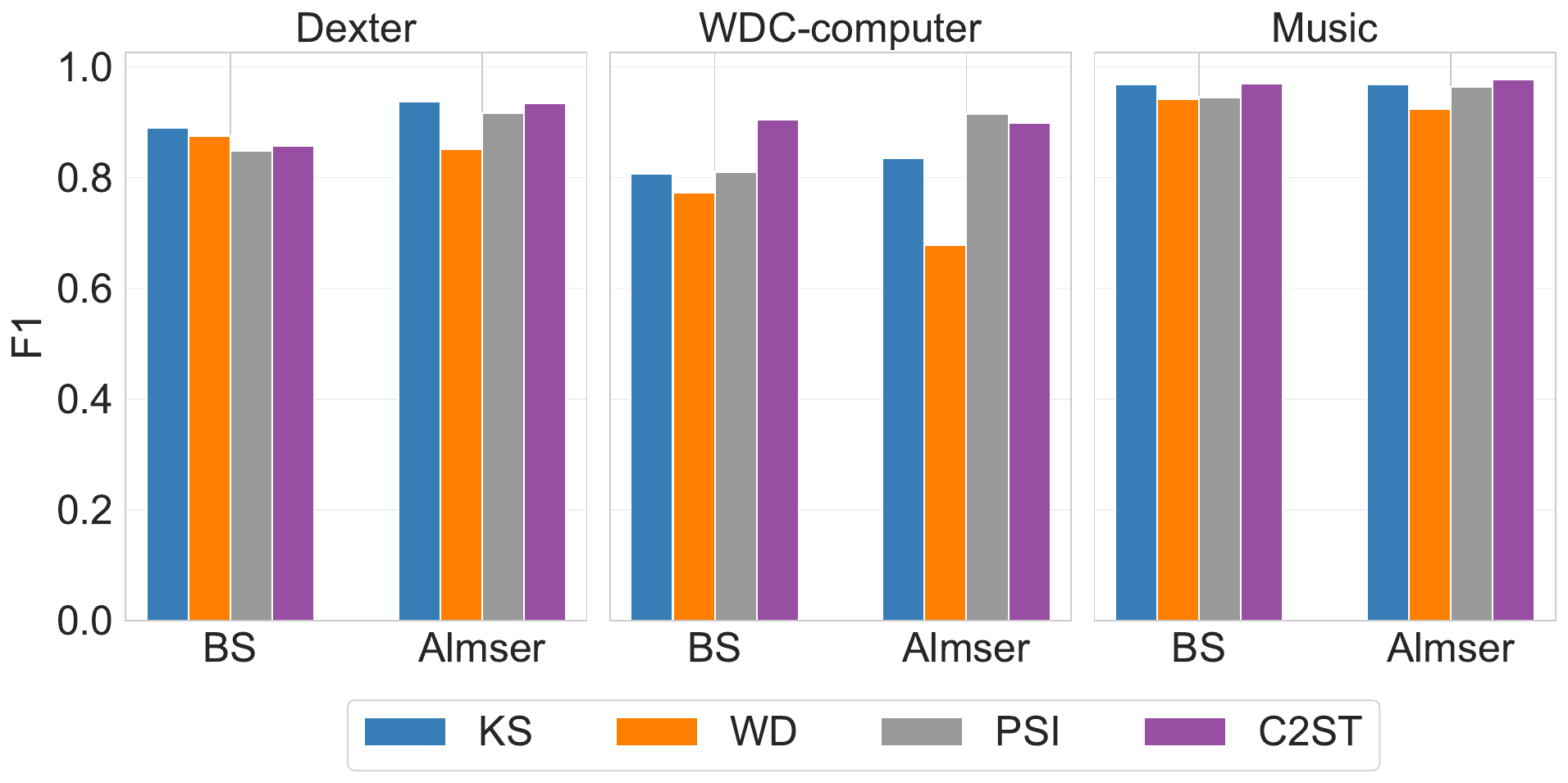}
        \caption{Budget = 1000}
        \label{fig:budget_1000}
    \end{subfigure}
    \hfill
    \begin{subfigure}[t]{5.6cm}
        \includegraphics[width=\linewidth]{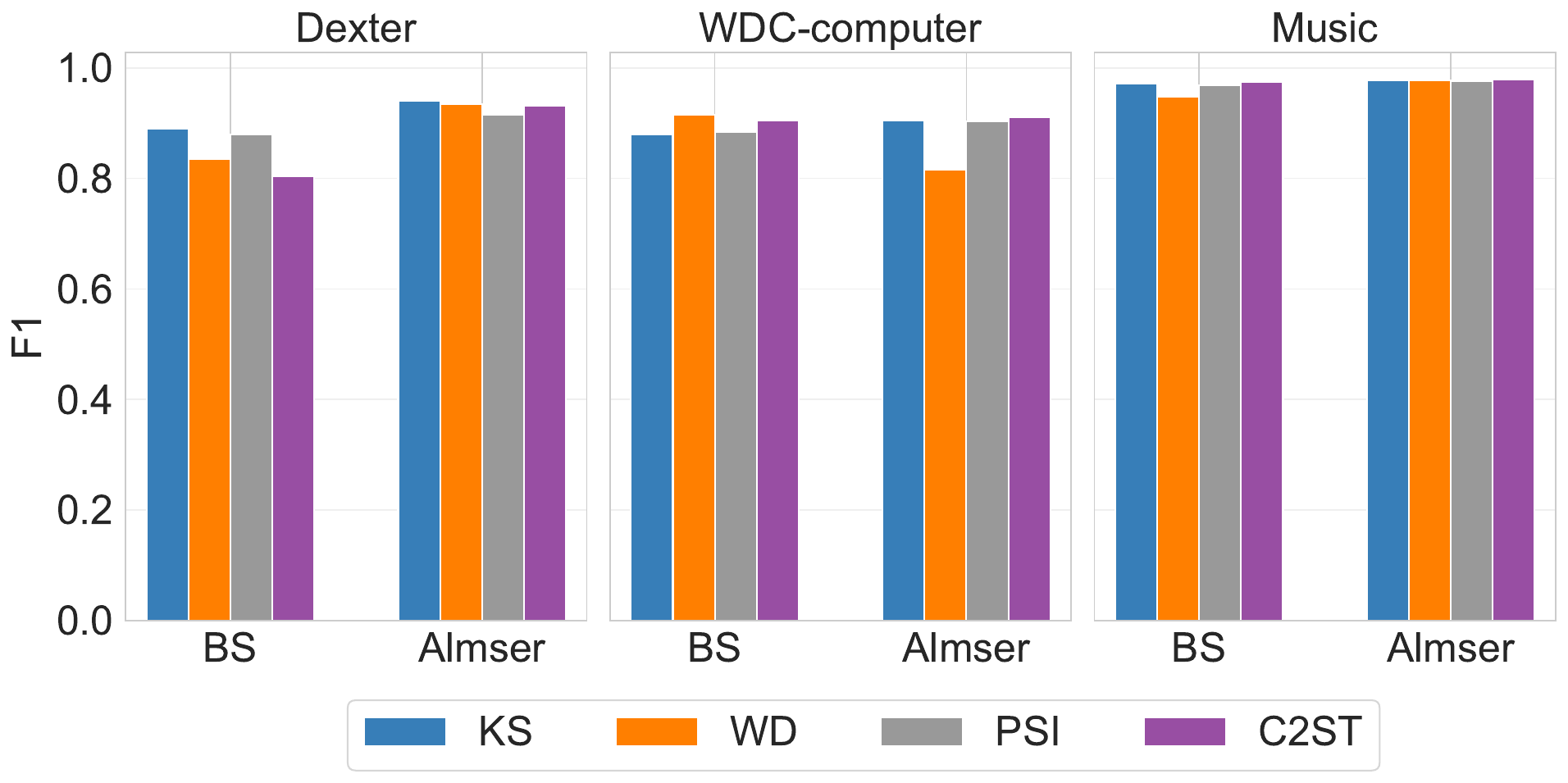}
        \caption{Budget = 1500}
        \label{fig:budget_1500}
    \end{subfigure}
    \hfill
    \begin{subfigure}[t]{5.6cm}
        \includegraphics[width=\linewidth]{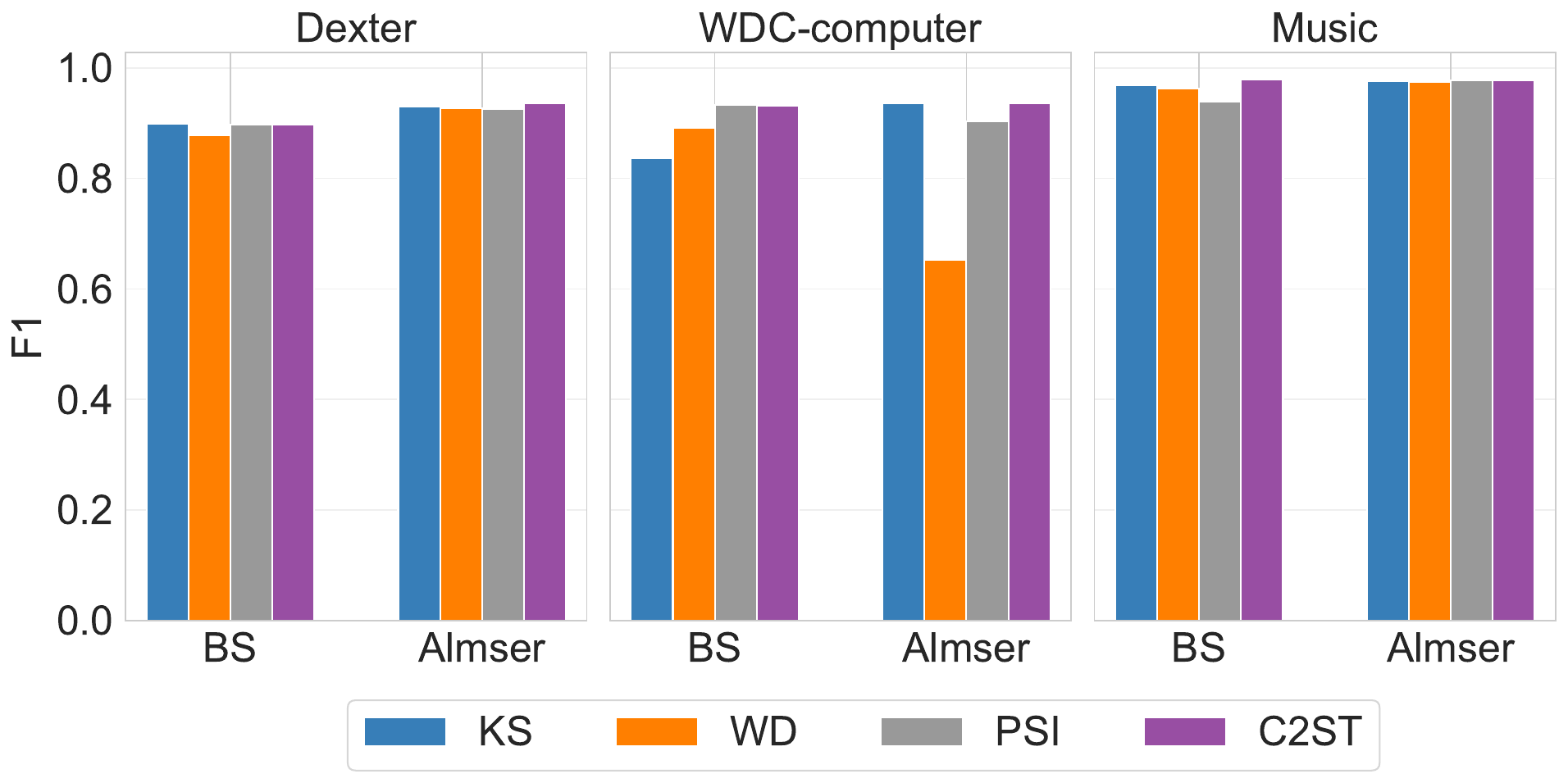}
        \caption{Budget = 2000}
        \label{fig:budget_2000}
    \end{subfigure}
    \caption{Comparison of the distribution tests using the \ac{al} methods \textit{Bootstrap} (BS) and \textit{Almser}.}
    \label{fig:stat_comparison}
\end{figure*}

\subsection{Analysis of Distribution Tests}
This section compares the various distribution tests for computing the similarity between \ac{er} problems. The similarity is essential for grouping tasks and determining which model $M_C^{i}$ of a cluster $C^i_{P}$ is used for a new \ac{er} problem $D_l\in\mathcal{N}$. \Cref{fig:stat_comparison} presents the results of the three distribution tests: Kolmogorov-Smirnov (KS), Wasserstein distance (WD), Population Stability Index (PSI), and the classifier two-sample test (C2ST). 

For the homogeneous \music dataset, all distribution tests performed similarly across both AL methods, indicating that the choice of distribution test is less critical for datasets with consistent characteristics. However, for heterogeneous or noisy datasets like \dexter and \wdc, the choice of distribution test and its combination with specific AL methods significantly impacts performance. 

For example, the Wasserstein Distance achieved strong results for the \wdc dataset when paired with the \textit{Bootstrap} method, but its performance was inconsistent for the \dexter dataset under the same method. This variability highlights the sensitivity of WD to dataset characteristics and AL configurations. In contrast, PSI demonstrated more robust and consistent results across different AL methods and datasets, making it a reliable choice for scenarios with diverse or noisy data. Notably, PSI achieved its best results for specific configurations, such as the \dexter dataset with a budget of 1000. 

The Kolmogorov-Smirnov test and {the multivariate analysis classifier two-sample test}, on the other hand, exhibited strong overall performance and proved to be reliable options across various datasets and AL methods. It slightly outperformed the other tests for the \dexter dataset, demonstrating its effectiveness in handling heterogeneous data.



\begin{figure}[t!]
      \begin{subfigure}[t]{4.7cm}
        \includegraphics[width=\linewidth]{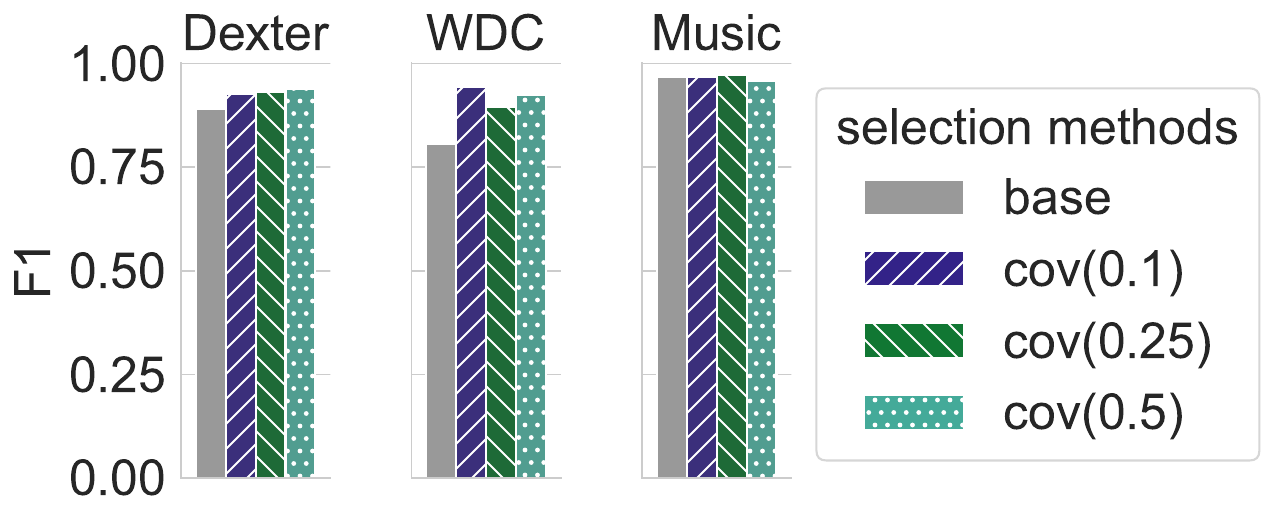}
        \caption{F1-Score results}
        \label{fig:recl_f1}
    \end{subfigure}
    \begin{subfigure}[t]{3.1cm}
        \includegraphics[width=\linewidth]{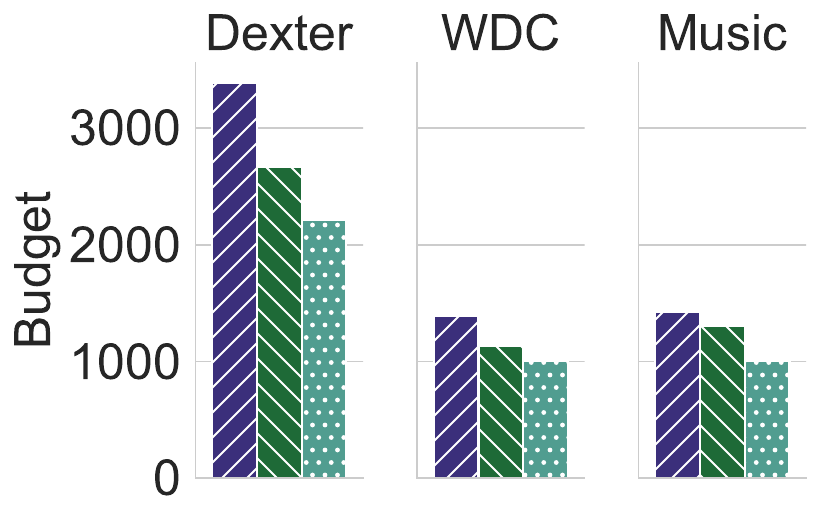}
        \caption{Labelling effort}
        \label{fig:recl_budget}
    \end{subfigure}
    \caption{Comparison of the selection strategies $sel_{base}$ and $sel_{cov}$ using Bootstrap \ac{al} (b=1000).}
    \label{fig:recluster_eval}
\end{figure}

\subsection{Analysis of Selection Methods}

We evaluate the impact of various selection methods for determining an appropriate model for a new \ac{er} problem, as described in \Cref{sec:new_d}. We use an initial budget of 1000 and the \textit{Bootstrap} \ac{al} method. The first method, denoted as $sel_{\text{base}}$, assigns an \ac{er} problem to the most similar cluster without any retraining. The second method, $sel_{\text{cov}}$, integrates the new \ac{er} problem into the \ac{er} problem graph $G_P$ and may update the existing models.

We analyze the ratios of uncovered similarity feature vectors during the training data generation step, specifically $t_{cov} = [0.1,0.25,0.5]$. For example, a ratio of 0.1 indicates that retraining with new training data occurs if a cluster contains at least 10\% of similarity feature vectors that were not included in the previous training data generation steps. ~\Cref{fig:recluster_eval} illustrates the results comparing the base selection strategy with the reclustering strategy in terms of resulting F1 scores and the increased labeling budget.

The reclustering selection strategy leads to improved \ac{er} quality compared to the base selection strategy. The $sel_{\text{cov}}$ achieves the best results for the \wdc and \music datasets, utilizing a threshold $t_{cov}$ of 0.1. However, these strategies also require the most additional labeling effort compared to the other selection methods. In the \dexter dataset, we observe a negative impact from a small covered ratio, such as 0.1. Due to the wide range of sizes of \ac{er} problems in the \dexter dataset, a single \ac{er} problem can trigger unnecessary retraining, as indicated by higher F1 scores when using a threshold $t_{cov}$ greater than 0.1.

\section{Discussion}\label{sec:discuss}
We introduce a novel method for reusing and creating classification models for integrated data sources under a limited labeling budget. This section discusses research questions, including limitations and suggestions regarding configurations and alternatives.

\smallskip
\textbf{RQ1: When should our method be applied?}
Our method performs best if more data sources are integrated, so a distinction between various \ac{er} problems and different classification models makes sense. For a small number of \ac{er} problems of moderate size, the resulting tasks can be solved using a single model determined by a sophisticated \ac{al} method such as \almser. However, due to the increasing runtime regarding a high number of integrated data sources, our method is more efficient and reaches similar or higher F1 scores compared to \almser. Transfer learning approaches are useful when it is known which source domain can be adapted to a certain target domain. Otherwise, the size of the training data from the source domain must be very large, as a small amount is not sufficient for transferring. \textcolor{black}{When \ac{er} problems originate from a general domain and sufficient training data is available or can be generated, \ditto yields qualitative results, particularly when leveraging data sources rich in textual information.}

\smallskip
{\textbf{RQ2: What are the additional costs of our method?} To reuse models for new \ac{er} tasks, we initially compute pairwise the similarities between the available \ac{er} tasks and cluster the resulting \ac{er} problem similarity graph. The similarity computation and the clustering scale with the number of initial \ac{er} tasks. However, the proportion of the total runtime is below 1\% with \almser and 70\% using \textit{Bootstrap}. Moreover, due to the reduced sets of record pairs for each cluster, the selection of informative record pairs using an \ac{al} method is more efficient than the application on the union of all \ac{er} tasks. In a real-world application, we would apply the retraining of clusters described in \Cref{sec:new_d} due to potential domain shifts. Consequently, we would run the clustering step for each new \ac{er} tasks, which is a moderate effort due to the efficiency of the Leiden algorithm~\cite{Traag2019leiden}. Moreover, we would retrain the underlying models with new selected training data from the upcoming \ac{er} tasks.}

\smallskip
\textbf{RQ3: Which active learning approach should be used?}
As the results indicate, our modified Bootstrapping outperforms the baseline methods and the combination with \almser regarding efficiency and achieves comparable results for clean \ac{er} problems. However, for dirty \ac{er} problems where data sources with duplicates are involved and a high number of problems, the combination with \almser should be used. 
\almser benefits from the reduced search space of candidate record pairs,  allowing \almser to focus on the most informative instances for labeling. The combination can be applied to heterogeneous and noisy datasets. If runtime efficiency is a critical requirement and the datasets are cleaner (e.g., \music dataset), the \textit{Bootstrap} AL method can be considered.  

\smallskip
\textbf{RQ4: Which distribution test for computing the similarity between \ac{er} problems is the most effective?}
The Wasserstein distance results in the lowest F1 score for the boot\-strapping-based \ac{al} approach on the \dexter and \music datasets. We hypothesize that the Wasserstein distance generates overly similar results for these datasets, which negatively affects performance, particularly in datasets with many \ac{er} tasks (\textit{Dexter}) or a large number of feature vectors (\textit{Music}). However, the Wasserstein distance outperforms the Kolmogorov-Smirnov test for the impure and smaller \wdc dataset, utilizing a certain budget. Meanwhile, the Population Stability Index achieves moderate linkage quality across the datasets for the \textit{Bootstrap} \ac{al} method.

We conclude that for homogeneous datasets like \music, the selection of a certain distribution test is less crucial than for more heterogeneous or noisy datasets like \dexter and \wdc. Moreover, the results suggest that Almser benefits more from specific distribution tests (e.g., Kolmogorov-Smirnov for \dexter), whereas \textit{Bootstrap} shows less variability across tests.

\smallskip
\textbf{RQ5: What are the limitations of our method?} 
Our method assumes the availability of a sufficient number of solved \ac{er} problems. However, when the number of solved problems is limited, the diversity of feature vectors becomes constrained. This limitation can result in models that fail to generalize effectively across the broader spectrum of \ac{er} problems, leading to a significant domain shift between solved and unsolved problems. 

To address this issue, we proposed selection strategies integrating the new \ac{er} problems in combination with an updating mechanism of the existing models. This process ensures that the models remain adaptable to evolving data distributions. In this work, we only consider the heterogeneity regarding the feature space. Therefore, we can only apply our method to \ac{er} problems with common features. A trivial strategy to handle data sources with different attributes is to apply a pre-trained language model to determine a comparable representation.

\section{Conclusion}\label{sec:concl}
The classification step in \ac{er} is crucial. The majority of existing approaches use ML models that require training data.  
Existing methods do not address the challenges regarding multiple classification tasks occurring in multi-source \ac{er} scenarios due to the increasing effort to generate training data. 
Our method addresses these issues by creating a model repository to reuse pre-trained classifiers for new \ac{er} problems. For initializing the repository, we cluster similar \ac{er} problems and derive one model for each group. We propose different variants to determine an appropriate model. In the base variant, we apply the most suitable built model of a similar cluster for a new \ac{er} task. The second selection method addresses potential domain shifts by updating the clusters and the models depending on the \ac{er} problems of a cluster. 

The evaluation demonstrates comparable quality to \almser and superior results to \transer, with significant runtime improvements,
highlighting the method's scalability and practical value. Compared to the language model-based approaches, \method achieves comparable performance or outperforms \ditto using 50\% of the training data, ranging from roughly 23K to 115K labeled record pairs. Comparing \method with the language model-based approaches \any, \unicorn, and \sudowoodo, our method significantly outperforms them.

In future work, we aim to integrate our method into a comprehensive system for storing and querying solved ER problems, clusters, and classification models, enabling solution reuse for new tasks. While our approach assumes a shared feature space, we suggested the combination of attributes into a single embedding using pre-trained language models. We also plan to explore incrementally expanding the feature space for specific ER tasks. {The effectiveness of the reuse depends on the clusters based on the \ac{er} problem graph and the similarities. As an additional criterion for the model performance of a cluster, we will investigate the relationship between model performance and cluster stability measures.}
\section*{Artifacts}
The reference code is available from our repository at \url{https://github.com/vicolinho/Model_Reuse_ER}.

\section*{Acknowledgment}
The authors acknowledge the financial support by the Federal Ministry of Education and Research of Germany and by Sächsische Staatsministerium für Wissenschaft, Kultur und Tourismus in the programme Center of Excellence for AI-research "Center for Scalable Data Analytics and Artificial Intelligence Dresden/Leipzig", project identification number: ScaDS.AI.

This work was partially funded by Universities Australia and the
German Academic Exchange Service (DAAD) grant 57701258.


\bibliographystyle{ACM-Reference-Format}
\bibliography{main}


\end{document}